\documentclass[10pt,a4paper]{article}
\usepackage[utf8]{inputenc}
\usepackage{amsmath}
\usepackage{amsfonts}
\usepackage{amssymb}
\usepackage{mathtools}
\usepackage{float}
\usepackage{tikz}
\usepackage{tikz-cd}
\usepackage{array}
\usepackage{adjustbox}
\usepackage{hyperref}
\usepackage{tcolorbox}

\usepackage{float}

\usepackage{ifpdf}
\usepackage{xcolor}
\usepackage{amsmath,amssymb,amsfonts,bm,amscd}
\usepackage{amscd}
\usepackage{mathtools}
\usepackage{graphicx, wrapfig}
\usepackage{multirow}
\usepackage{verbatim}
\usepackage{appendix}
\usepackage{fancybox}
\usepackage{slashed}
\usepackage{extarrows}
\usepackage{braket}
\usepackage{url}
\usepackage[backend=bibtex8]{biblatex}

\usepackage{bbold}

\newcommand{\bZ}{\mathbb{Z}}
\newcommand{\bC}{\mathbb{C}}

\newcommand{\bR}{\mathbb{R}}

\newcommand{\id}{\text{id}}

\usepackage{tikz}
\usetikzlibrary{decorations.markings}
\usetikzlibrary{decorations.pathreplacing,angles,quotes}
\usetikzlibrary{decorations.pathmorphing,shapes}
\newcounter{sarrow}

\definecolor{WScolor}{RGB}{191,191,255}
\definecolor{WScolor light}{RGB}{224,224,255}
\definecolor{DefectColor}{RGB}{0,0,128}
\tikzset{
	defect/.style={color=DefectColor, line width=1.5},
	arrow position/.style={postaction={decorate,decoration={
		markings,
		mark=at position #1 with {\arrow{>}}
	}}},
	opp arrow position/.style={postaction={decorate,decoration={
		markings,
		mark=at position #1 with {\arrow{<}}
	}}},
	defect node/.style={circle,inner sep=1.5pt,fill, DefectColor},
	snake it/.style={
		-stealth,
		decoration={
			snake, 
    		amplitude = .4mm,
    		segment length = 2mm,
    		post length=0.9mm
    	},
    	decorate
    }
}

\usetikzlibrary{cd}
\usepackage[margin=2.7cm]{geometry}

\newcommand{\iprod}{\mathbin{\lrcorner}}

\begin{document}

\title{Defects and Phases of Higher Rank Abelian GLSMs
}

\author{%
	Ilka Brunner$^*$  \quad Daniel Roggenkamp$^{\vee\#}$ \quad Christian P.~M.~Schneider$^*$
	\\[0.5cm]
	\normalsize{\texttt{\href{mailto:ilka.brunner@physik.uni-muenchen.de}{ilka.brunner@physik.uni-muenchen.de}}} \\ 
	\normalsize{\texttt{\href{mailto:daniel.roggenkamp@mis.mpg.de}{daniel.roggenkamp@mis.mpg.de}}}\\%
	\normalsize{\texttt{\href{christian.schneider@physik.uni-muenchen.de}{christian.schneider@physik.uni-muenchen.de}}} 
		\\[0.3cm]  %
	\hspace{-1.2cm} {\normalsize\slshape $^*$Arnold Sommerfeld Center,  LMU M\"unchen, Theresienstra\ss e 37, 80333 M\"unchen, Deutschland}\\[-0.1cm]
		\hspace{-1.2cm} {\normalsize\slshape $^\vee$Max Planck Institute for Mathematics in the Sciences, Inselstra{\ss}e 22, 04103 Leipzig, Deutschland}\\[-0.1cm]
		\hspace{-1.2cm} {\normalsize\slshape $^\#$
		Institute for Theoretical Physics, University of Leipzig,
Br\"uderstra{\ss}e 16, 04103 Leipzig, Deutschland}
}

\date{}
\maketitle

\begin{abstract} 

We construct defects describing the transition between different phases of gauged linear sigma models with higher rank abelian gauge groups, as well as defects embedding these phases into the GLSMs. Our construction refers entirely to the sector protected by B-type supersymmetry, decoupling the gauge sector. It relies on an abstract characterization of such transition defects and does not 
involve an actual perturbative analysis. 
It turns out that the choices that are required to characterize consistent transition defects match with the homotopy classes of paths between different phases.

Our method applies to non-anomalous as well as anomalous GLSMs, and we illustrate both cases with examples. This includes the GLSM associated to the resolution of the $A_N$ singularity and one describing the entire parameter space of $N=2$ minimal models, in particular, the relevant flows between them.

Via fusion with boundary conditions, the defects we construct yield functors describing the transport of D-branes on parameter space. We find that our results match with known results on D-brane transport.

\end{abstract}

\thispagestyle{empty}

\vfill
LMU-ASC 20/24

\newpage

\section{Introduction}

In this paper, we consider abelian gauged linear sigma models (GLSMs) in 2 dimensions with $(2,2)$ supersymmetry \cite{Witten:1993yc}. Such theories are specified by an abelian gauge group $U(1)^n$, a collection of matter multiplets and, possibly, a superpotential compatible with the gauge symmetry.  
 In the non-anomalous Calabi-Yau case, such gauge theories exhibit a moduli space, parametrized by the (complexified) Fayet-Iliopoulos parameter, subject to D- and F-term constraints. 

In general, GLSMs exhibit different phases, characterized by (partial) breaking of the gauge symmetry. While a discrete gauge symmetry remains unbroken in an orbifold phase, the gauge symmetry is completely broken in geometric phases. After quantum corrections have been taken into account, the moduli space features singular codimension $2$ loci, and different phases can be connected by homotopically different paths avoiding the singularities.

In this paper, we will utilize the idea that defects can  relate the physics at different points in moduli space, see \cite{Brunner:2007ur,douglas:2010ic,Klos:2020upw,gaiotto:2015aoa,Galakhov:2020upa}. Such  defects are codimension one domain walls, separating two theories at different points in moduli space.  One can think of  them as originating from a trivial defect line in one of the theories. Upon perturbing the theory at one side of the defect, one obtains a defect between initial and perturbed theory.  As this defect depends on the specific perturbation, one expects different defects for different paths in moduli space.

We  restrict to a sector of GLSMs protected by B-type supersymmetry and construct B-type defects that connect different phases, as well as defects that mediate between orbifold phases and the GLSM. Such defects become topological upon B-twist, and hence their worldsheet support can be deformed in arbitrary ways. Within the SUSY protected sector, they can for example wrap operator insertions, mapping the bulk operators of one phase to another one, respectively to the GLSM, or they can be merged with boundaries, relating the respective categories of D-branes.

Indeed, part of the data required to specify D-branes consists of a choice of representation on the Chan-Paton degrees of freedom. Going from the GLSM with a continuous gauge symmetry $U(1)^n$ to an orbifold phase, the representation of the initial gauge group simply restricts to a representation of the subgroup. In the other direction, a lift from an orbifold phase to the GLSM involves a choice of lift of the representation. A priori, there are many different possibilities. In the case that the lift of the D-brane to the GLSM can be completed by a flow to another phase, describing a path in moduli space from one phase to another, there are however restrictions on the possible lifts \cite{herbst2008}.

In the approach we follow in this paper, these consistent lifts are implemented in a manifestly functorial way by defects. On general grounds, lift and deformation defects have some defining properties \cite{Klos:2020upw}, among them is semi-invertibility. This implies that to any such defect one can associate a projector that singles out a subcategory of the GLSM D-branes. We propose a concrete construction that yields defects that satisfy all expected properties. The starting point   is the identity defect of the GLSM which  is $U(1)^n\times U(1)^n$ equivariant. Pushing the theory on one side of this defect into an orbifold phase breaks the symmetry to $U(1)^n \times G$, where $G$ is the unbroken gauge symmetry in the orbifold phase. The Chan-Paton representations on the defect transform to the respective induced representations of the broken subgroup. This however is not the whole story: We find, that in order to obtain consistent defects, we have to impose additional truncations of the representations. The truncation depends on the path of deformation in the GLSM parameter space. For each phase boundary to be crossed, we introduce an upper bound on the charges under the distinguished $U(1)$ gauge symmetry preserved on this phase boundary. The choice of upper bounds, which we call `cutoff' parameters characterizes our defect completely. It specifies to which subcategory of GLSM branes any brane of the phase can be lifted.

The question how D-branes are transported in gauged linear sigma models was previously addressed in the program initiated in \cite{herbst2008}.
Based on an analysis of the gauge sector, one of the proposals put forward in \cite{herbst2008} is the `band restriction rule' for higher rank Abelian gauge theories,
as is first explained from the perspective of a boundary potential in \cite{herbst2008}. It states  that for a smooth brane transport from an orbifold phase to a (partially) resolved phase, there is a restriction on the possible lifts from the phase to the GLSM, singling out subcategories of the GLSM. The possible choices compatible with the band restriction rule  correspond to the different homotopy classes of paths in moduli space.
In subsequent work, the result was confirmed by arguments using analyticity  of the hemisphere partition function \cite{Hori:2013ika}, extended to the non-Calabi-Yau case in \cite{Clingempeel:2018iub}, to hybrid models in \cite{Knapp:2024wwh}, see also \cite{Lin:2024fpz,Knapp:2023izn} for some results on the non-Abelian case.\footnote{The grade restriction rule inspired mathematicians to
 construct equivalences between D-brane categories, see \cite{Segal:2010cz,Donovan:2013gia}.}
Our results are in complete agreement with the band restriction rule, where our choice of cutoff parameters precisely match with the bands appearing in \cite{herbst2008}.

Since our construction is entirely in the sector protected by B-type supersymmetry, the defect constructions apply in both anomalous as well as non-anomalous cases.

We apply our construction to a variety of different models. First of all, we consider orbifolds  ${\mathbb C}^2/{\mathbb Z}_N$, which are orbifold phases of GLSM with $U(1)^{N-1}$ gauge group. Our construction of a defect between the orbifold and GLSM provides a concrete functor that implements the MacKay correspondence between the orbifold category and resolved phases. We furthermore consider a two-parameter Calabi-Yau model with a Landau-Ginzburg phase, a geometrical phase and 2 mixed phases. In both examples we compare our results to the band restriction rule and find complete agreement. 

One of the advantages of our construction is that it does not rely on the Calabi-Yau condition and can be used also for relevant flows. To exemplify this, we consider $N=2$ minimal models, described in terms of Landau-Ginzburg orbifolds with superpotential $W=X^d$ and orbifold group ${\mathbb Z}_d$. 
Such models $X^d/{\mathbb Z}_d$ with different $d$ are connected by relevant flows. The latter are non-perturbative, hence, it is particularly useful to understand them from a defect perspective. A direct construction was given in \cite{Brunner:2007ur} and shown to be mirror to the flows considered in \cite{Hori:2000ck}. As it turns out, these flows can be modelled in a GLSM and are hence amenable to the approach of this paper. Namely, we show that all the Landau-Ginzburg orbifold models $X^i/{\mathbb Z}_i$, $i\leq d$ arise as different phases of a single GLSM with gauge group $U(1)^{d-2}$. 
Applying our proposed GLSM construction in this setup, we find all flow defects from \cite{Brunner:2007ur}.

This paper is organized as follows: In Section~\ref{sec:outline}, we outline the general idea to associate defects to perturbations of quantum field theories. In Section~\ref{sec:Review} we turn to the concrete case of gauged linear sigma models. We review the relevant description of B-type defects in GLSMs and Landau-Ginzburg orbifolds in terms of equivariant matrix factorizations. Particular emphasis is put on the identity defect of GLSMs, as it provides the starting point for our construction of lift and flow defects. 
In Section~\ref{sec:examples} we illustrate our approach in concrete examples of non-anomalous GLSMs with orbifold as well as geometric phases. Section~\ref{sec:minmods} is devoted to an anomalous example -- the GLSMs containing the flows between $N=2$ minimal models.

\section{Defects, Flows and Deformations}\label{sec:outline}

Quite often, in physics we do not consider isolated quantum field theories, but families which are fibered over a moduli space ${\cal M}$. For each $p\in {\cal M}$ there is a quantum field theory $T(p)$, whose  massless physical fields correspond to local deformations. Physical quantities, such as correlation functions or possible boundary conditions, in general depend on the point in moduli space. To move from one point to another, one can invoke perturbation theory and derive the physics at a nearby point. 
However, as is well known, moduli spaces may in general have a non-trivial geometry. For instance there might be several homotopically different paths connecting the same pair of points, and transport of quantities along those paths might lead to different results.  

To approach questions about the global structure of moduli spaces, one would like to invoke an approach that is non-perturbative in the coupling constants. Here, defects come into play. To explain the basic idea, consider an initial theory $T(p)$ associated to a point $p$ in moduli space on some space time manifold $X$. Choose a codimension $1$ submanifold of $X$, which separates space-time into two disjoint regions. Insert a domain wall (or defect) along this submanifold which glues together the theories on either side of it in a trivial way. Because of the trivial gluing, the domain wall is invisible, i.e.~it does not affect the correlation functions, but is merely a formal trick which allows for the separate treatment of the quantum field theories on the two domains. 
Then perturb the theory, but only in the region on one side of the domain wall. 
This deforms the theory in that region to a new point $q$ in the moduli space, whereas the theory on the other side of the domain wall remains at $p$. 
The domain wall has to go along in the process and is deformed from the invisible domain wall in theory $T(p)$ to a non-trivial one separating the original theory $T(p)$ from the deformed theory $T(q)$. 
This is procedure is depicted in \eqref{pic:flow} below.

\begin{equation}\label{pic:flow} 
\tikz[baseline=10]{
	\fill[WScolor] (-2,0) rectangle (4,1);
	\draw[densely dashed] (1,0) -- (1,1);
	\node at (1.55,.2) {$I_\text{T(p)}$};
	\node at (-.5,.5) {T(p)};
	\node at (2.5,.5) {T(p)};
}\;\tikz[baseline=-4]{
\node at (0.7,0.4) {perturbation};
	\path (0,0) edge[snake it] (1.5,0);
}\;\tikz[baseline=10]{
	\fill[WScolor light] (-2,0) rectangle (1,1);
	\fill[WScolor] (1,0) rectangle (4,1);
	\draw[defect] (1,0) -- (1,1);
	\node at (-.5,.5) {T(q)};
	\node at (2.5,.5) {T(p)};
	\node at (1.3,.2) {$R$};
}
\end{equation}

The domain wall between $T(p)$ and $T(q)$ created in this way is called a deformation defect. It captures all the information about the transport from theory $T(p)$ to $T(q)$ along the chosen path of deformation. 

For nearby points $p$ and $q$ perturbation theory can be used to determine deformation defects, and their properties, such as their reflectivity and transmissivity \cite{Brunner:2015vva}. 
For finite distances, explicit computations can be hard, not least because even exactly marginal deformations of the bulk theory can trigger relevant renormalization group flows on the defect. Still, however difficult to compute in practice, 
the resulting defect  connects the theories $T(p)$ and $T(q)$. A priori it depends on the chosen path of deformation from $p$ to $q$ -- two different paths between $p$ and $q$ can lead to different deformation defects between $T(p)$ and $T(q)$. 

\subsection{Perturbation Defects in Supersymmetric Theories}

To make efficient use of the defect point of view, we turn to a setting where  defects are under good control, namely to topological quantum field theories. These theories allow for an efficient description of defects in terms of 1-morphisms in 2-categories. Due to the absence of singularities in this setting defects can be merged. This leads to a well defined fusion product of defects, which is captured by the composition of 1-morphisms.

The fusion product is important for the study of moduli spaces, because fusion with deformation defects encodes the behavior of defects under deformations of the underlying bulk theory. 
For instance, the figure in \eqref{pic:boundarydef} below illustrates the action of deformation defects on boundary conditions\footnote{Boundary conditions in a quantum field theory can be considered as defects between this and the trivial quantum field theory.}, which captures the behavior of the latter under the respective deformation of the bulk theory. Note that since fusion of defects is functorial, 
the defect approach to bulk deformations immediately implies functoriality of the 
behavior of boundary conditions under bulk deformations. 
\begin{equation}\label{pic:boundarydef} 
\begin{aligned}
\tikz[baseline=17,scale=0.8]{
		\fill[WScolor] (-1,.25) rectangle (2,1.25);
		\draw[defect] (2,.25) -- (2,1.25);
		\node at (1.7,.45) {$B$};
		\node at (.5,.75) {T(p)};
	}\quad&\longmapsto\quad\tikz[baseline=17, scale=0.8]{
		\fill[WScolor light] (-2,.25) rectangle (1,1.25);
		\fill[WScolor] (1,.25) rectangle (4,1.25);
		\draw[defect] (1,.25) -- (1,1.25);
		\draw[defect] (4,.25) -- (4,1.25);
		\node at (.7,.45) {$D$};
		\node at (-.5,.75) {T(q)};
		\node at (2.5,.75) {T(p)};
		\node at (3.7,.45) {$B$};
	}\phantom{*}&\tikz[baseline=-2]{
	\path (0,0) edge[snake it] (0.7,0);
}\;\phantom{**}
	\tikz[baseline=17,scale=0.8]{
		\fill[WScolor light] (-1,.25) rectangle (2.5,1.25);
		\draw[defect] (2.5,.25) -- (2.5,1.25);
		\node at (1.7,.45) {$D\otimes B$};
		\node at (0.4,.75) {T(q)};
	}
\end{aligned}
\end{equation}
Moreover, also the concatenation of deformations is encoded in the fusion of the respective deformation defects. 

The topological quantum field theories we would like to consider arise as topological twists of  supersymmetric theories. Perturbations of these theories 
compatible with the topological twist can be pushed to the level of the twisted theory and become deformations of the latter. 

Specifically we are interested in two-dimensional 
 $N=(2,2)$ superconformal field theories. These theories feature two types of topological twists, the A-twist and the B-twist, which require A-type and B-type supersymmetry, respectively. 
Perturbations of such a theory originating from its $(c,c)$ ring preserve A-type supersymmetry and are hence compatible with the A-twist. The respective deformation defects preserve A-type supersymmetry. Likewise, perturbations originating from the $(a,c)$ ring preserve B-type supersymmetry, are compatible with the B-twist and lead to B-type supersymmetric deformation defects \cite{Brunner:2007qu}.
 
The fusion of defects of the same type is protected by supersymmetry. This includes fusion of defects with boundary conditions. Indeed, in the context of bulk perturbations of boundary conditions this is expected, since
A-, respectively B-type D-branes remain supersymmetric under (c,c), respectively (a,c) perturbations. Note however that conformal symmetry at the boundary may well be broken by such a deformation, inducing relevant flows of the D-branes to new stable supersymmetry preserving configurations \cite{Brunner:2009mn,Bachas:2013nxa}. 
In a situation where the boundary condition has a space time interpretation as a BPS particle, this has an interpretation as the appearance of a line of marginal stability.
However, it is not necessary to refer to space time, and one can encounter relevant flows for defects as well as for boundary conditions. Away from these lines of marginal stability, defects can be minimally adjusted to the perturbed bulk-background. 

On the level of the twisted theory, the D-brane category `decouples' from the respective bulk moduli in the sense that the category of D-branes is the same for each point in moduli space. For instance in the case of B-twisted non-linear sigma models, 
this is the statement that the category of coherent sheaves on the target space is independent of its K\"ahler moduli. However, crossing lines of marginal stability leads to a non-trivial action of autoequivalences of these categories. In particular, monodromies of the moduli space manifest themselves in terms of autoequivalences of the D-brane category \cite{Aspinwall:2001dz}. Indeed, these autoequivalences are special instances of deformation defects \cite{Brunner:2008fa}. 

 While in the above discussion we pretended that the deformations of the bulk theory are marginal, our arguments also apply  to relevant deformations: in the figure in \eqref{pic:flow}, we may turn on a relevant instead of an exactly marginal perturbation, inducing a non-trivial RG flow of the theory on the left side of the defect. The resulting flow defect connects an initial UV theory with an IR theory. On the level of the topologically twisted theory, there is no difference between relevant and marginal deformations of the initial, untwisted theory. 

\subsection{Properties of Deformation Defects}

Deformation defects are in general difficult to construct by explicitly following the combined bulk-defect flow. It is therefore of interest to characterize them abstractly. Indeed they have characteristic properties with respect to fusion. 
For instance, one may imagine to perturb a bulk theory
everywhere, except in some bounded region of space time, creating a deformation defect on the boundary of this domain. Shrinking the region to zero size, the deformation is just extended to all of space time. The result is the deformed theory with no defect insertion as depicted in \eqref{pic:transfusion}. 

\begin{equation}\label{pic:transfusion}
 \tikz[baseline=20,scale=0.7]{
		\fill[WScolor light] (0,0) rectangle (2,1.5);
		\fill[WScolor] (2,0) rectangle (4,1.5);
		\fill[WScolor light] (4,0) rectangle (6,1.5);
		\draw[defect] (2,0) -- (2,1.5);
		\draw[defect] (4,0) -- (4,1.5);
		\node at (3,.75) {$T(p)$};
		\node at (1,.75) {$T(q)$};
		\node at (5,.75) {$T(q)$};
		\node at (2.1,-.5) {$R^{pq}$};
		\node at (4.1,-.5) {$T^{pq}$}
	}\tikz[baseline=6]{
	\path (0,0) edge[snake it] (0.7,0);
}
\tikz[baseline=20,scale=0.7]{
		\fill[WScolor light] (-1,0) rectangle (2,1.5);
		\fill[WScolor light] (2,0) rectangle (5,1.5);
		\draw[defect] (2,0) -- (2,1.5);
		\node at (3.5,.75) {$T(q)$};
		\node at (0.5,.75) {$T(q)$};
				\node at (2.0,-.5) {$R^{pq}\otimes T^{pq}$};
	}
	\tikz[baseline=6]{
	\node at (0,0) {$\cong$};}
	\tikz[baseline=20,scale=0.7]{
		\fill[WScolor light] (-1,0) rectangle (2,1.5);
		\fill[WScolor light] (2,0) rectangle (5,1.5);
		\draw[densely dashed] (2,0) -- (2,1.5);
		\node at (3.5,.75) {$T(q)$};
		\node at (0.5,.75) {$T(q)$};
				\node at (2.2,-.5) {$I_{T(q)}$};
	}
\end{equation}

This means that fusing the deformation defect $R^{pq}:T(p) \to T(q)$ and its adjoint $T^{pq}: T(q)\to T(p)$ yields the identity defect in the deformed theory.

\begin{equation}
\label{eq:RotimesT}
R^{pq}\otimes T^{pq} = \id_{T(q)}
\end{equation} 

This argument does not apply in the opposite direction: shrinking the region of deformation to zero size does not result in just the undeformed theory, at least for relevant perturbations. Relevant perturbations reduce the degrees of freedom, and so shrinking the perturbation region leaves behind a defect which projects onto the IR degrees of freedom in the UV theory. Indeed, as a direct consequence of equation \eqref{eq:RotimesT} one finds 
\begin{equation}
(T^{pq}\otimes R^{pq})^2 = T^{pq}\otimes R^{pq} \otimes T^{pq}\otimes R^{pq} = T^{pq}\otimes R^{pq}
\end{equation}
Hence, the defect $P^{pq}:=T^{pq}\otimes R^{pq}$ is idempotent with respect to defect fusion. It indeed projects on IR degrees of freedom.

While deformation defects have some more special properties \cite{Klos:2020upw}, we will only use the ones associated with fusion here. 

\subsection{Applications to Abelian Gauged Linear Sigma Models}

Gauged linear sigma models provide  UV descriptions of IR  models. In different limits of the GLSM parameters, phases of different nature can arise. In the non-anomalous Calabi-Yau case, the family of UV theories flows to a family of IR theories. Hence the GLSM parameters
describe an honest family of IR quantum field theories. 
In the non-Calabi-Yau case, the GLSM parameters are not exactly marginal but relevant and therefore trigger non-trivial flows in the IR theories. Hence one 
distinguishes UV and IR phases of the GLSM which are connected by relevant flows. The techniques we discuss in subsequent chapters apply to both situations.

Any two different phases (or points of the IR moduli space) $i$ and $j$ are connected by  bulk deformations. As outlined above, to any deformation path $\gamma$ we can associate a deformation defect
\begin{equation}
R^{ij}_\gamma : T(i) \to T(j)
\end{equation}
In general, there may be different paths, and, as discussed above, we expect the deformation defects to depend on it. Note that this in particular applies to monodromies.

As the GLSM provides a UV description of the IR physics, one may ask whether one can realize a given path in the IR moduli space, and with it the corresponding defects, by taking a ``detour" via the GLSM. Then the deformation defects $R^{ij}_\gamma$ between phases of the GLSM $R^{ij}_\gamma$ should decompose as
\begin{equation}
R^{ij}_\gamma= R^i_\gamma \otimes T^j_\gamma
\end{equation}
Here, $R^i_\gamma: GLSM\to i$ and $T^j_\gamma: j \to GLSM$ are defects from the GLSM to phase $i$ and from phase $j$ to the GLSM which should correspond to 
flow defects from the GLSM to the phases, respectively their duals.
Therefore, they should satisfy equation \eqref{eq:RotimesT}:
\begin{equation}
R^i_\gamma \otimes T^i_\gamma = \id_i
\end{equation}
where, $\id_i$ is the identity defect of the theory in phase $i$. Moreover the defects
\begin{equation}
P^i_\gamma:=T^i_\gamma \otimes R^i_\gamma :\text{GLSM}\to \text{GLSM}
\end{equation}
are defects in the GLSM which project on the degrees of freedom of phase $i$. 
By their actions on boundary conditions, all the defects mentioned here give rise to functors between the categories of D-branes of the respective theories. For instance, the $T^i_\gamma$ embed the D-brane categories of phase $i$ into the GLSM, and the $P^i_\gamma$ project onto the subcategory of GLSM branes which correspond to the D-branes in phase $i$. 
All of this is collected in the following diagram, taken from \cite{Brunner:2021cga}.

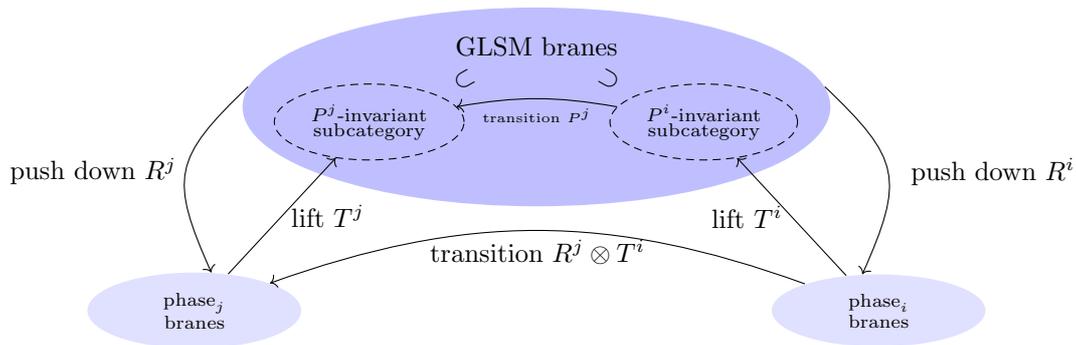
\[\tikz{
\node[ellipse, fill=WScolor,minimum width=220,minimum height=75](A) at (0,-.8){};
	\node (F) at (0,0) {GLSM branes};
	\node[ellipse, fill=WScolor, draw, style=densely dashed] (B) at (-2.2,-1) {$\substack{P^j\text{-invariant}\\\text{subcategory}}$};
	\node[ellipse, fill=WScolor, draw, style=densely dashed] (C) at (2.2,-1) {$\substack{P^i\text{-invariant}\\\text{subcategory}}$};
	\node[ellipse, fill=WScolor light,minimum width =80] (D) at (-4.5,-3.5) {$\substack{\text{phase}_j\\\text{branes}}$};
	\node[ellipse, fill=WScolor light,minimum width =80] (E) at (4.5,-3.5) {$\substack{\text{phase}_i\\\text{branes}}$};
	
	\path (A) edge[bend left=35,looseness=1.4,->] node[right]{\phantom{*}push down $R^i$} (E);
	\path (A) edge[bend right=35,looseness=1.4,->] node[left]{push down $R^j$} (D);
	\path (B) edge[<-, bend left=10] node[below]{\tiny transition $P^j$} (C);
	\path (D) edge[->] node[right]{lift $T^j$} (B);
	\path (E) edge[->] node[left]{lift $T^i$} (C);
	\path (D) edge[<-, bend left=20] node[below]{transition $R^j\otimes T^i$} (E);
	\path (B) -- node[rotate=30]{$\subset$} (F);
	\path (C) -- node[rotate=150]{$\subset$} (F);
}\]

In this paper, we focus on the defects $T^i_\gamma$ embedding orbifold phases into GLSMs. 
More precisely, we propose a concrete construction of these defects for GLSMs with higher rank abelian gauge groups, which generalizes the construction given in \cite{Brunner:2021cga} for the rank-one case. Central part of the construction is a prescription for how defects in abelian GLSMs 
behave when the GLSM on one side of it is pushed to an orbifold phase. Applied to the identity defect in the GLSM this yields the defects $T^i_\gamma$. We will spell out this construction in 
Section~\ref{sec:Review} and then apply it to concrete examples in Sections~\ref{sec:examples} and \ref{sec:minmods}.
\section{Defects and Phase Transitions in GLSMs} \label{sec:Review}

In the previous section we described some features of defects associated to deformations or perturbations of bulk theories. In this section we will lay out a procedure to construct such flow defects in the context of gauged linear sigma models with abelian gauge groups. It generalizes a construction for rank-one abelian gauge groups given in \cite{Brunner:2021cga}. 
This procedure allows for a very explicit non-perturbative construction of defects describing phase transitions between phases of GLSMs. In Sections~\ref{sec:examples} and \ref{sec:minmods} below we will illustrate it in conrete examples.
We will start this section with a brief overview of GLSMs, their phases and defects in these models.

\subsection{The GLSMs}\label{sec:GLSM}

We consider $\mathcal{N}=(2,2)$ gauged linear sigma models in $1+1$ dimensions with abelian gauge groups \cite{Witten:1993yc}. 
Such theories are specified by tuples
\begin{equation}
(G,V,(r,\theta),W),
\end{equation}
where $G$ is a compact abelian Lie group, $\rho:G\rightarrow GL(V)$ is a faithful unitary representation of $G$, $(r,\theta)\in\mathbb{R}^{\dim(G)}\times\mathbb{R}^{\dim(G)}$ are real parameters such that $\exp(r_j+i\theta_j)\in\text{Hom}(\pi_1(G),\mathbb{C}^*)^{\text{Ad}_G}$ and ${W\in\text{Sym}(V^\vee)^G}$ is a $G$-invariant polynomial.

In terms of physics, this data has the following interpretation: $G$ is the gauge group, $V$ describes the chiral matter content, and $W$ is the superpotential. $(r,\theta)$ are the Fayet-Iliopoulos- (FI) and $\theta$-parameters determining the IR-behavior of the theory. In the following, we will restrict our attention to the case
$G=U(1)^n$ and $V=\mathbb{C}^k$. We also assume that there is a vector R-symmetry $R:U(1)_V\rightarrow GL(V)$ such that the superpotential has R-charge $2$.

The representation $\rho$ will be specified by a charge matrix $Q_{ai}$, where $a\in \{1, \dots, n\}$ is a gauge index and $i\in \{1, \dots , k \}$ is a flavor index. In particular, there are vectors $Q_i:=(Q_{a\,i})_{a=1}^n\in \bR^n$ associated to each flavor multiplet. 

The IR behavior of this model is captured by its scalar potential 
\begin{equation}
U=\sum_{i=1}^k\left\vert\sum_{a=1}^n Q_{ai} \sigma_a X_i\right\vert^2+\frac{e^2}{2}\sum_{a=1}^n\left(\sum_{i=1}^k Q_{ai} \vert X_i\vert^2-r_a\right)^2+\sum_{i=1}^k\left\vert\frac{\partial W}{\partial X_i}\right\vert^2.
\end{equation} 
Here $\sigma$ is the scalar component of the vector multiplet -- which we later on assume to be decoupled -- and the $X_i$ are the chiral superfields. The vanishing locus ${\mathcal M}:=\{U=0\}/G$ of $U$ modulo gauge transformations is the vacuum manifold of the model. 

The IR behavior crucially depends on the FI parameters. In fact, there are domains in the parameter space where the theory exhibits qualitatively different IR descriptions. These are called the phases of the GLSM. 
There are different kinds of phases distinguished among other things by the amount of gauge symmetry breaking. So-called geometric phases for instance are characterized by a complete breaking of the gauge symmetry. The modes transverse to the vacuum manifold are massive and the IR in such a phase is described by a sigma model on ${\mathcal M}$. In Landau-Ginzburg phases on the other hand the vacuum manifold only consists of a single point, with massless transverse modes. The gauge symmetry is broken to a finite subgroup, and the low energy theory is described by an orbifold of a Landau-Ginzburg model. Beyond these two there are various types of mixed phases.

The boundaries between different phases in parameter space can be described by positive cones of charge vectors: For any $I\subset\{1,\ldots,k\}$ we can associate the positive cone
\begin{equation}\label{eq:cone}
{\mathrm{Cone}}_I = \left\{ \sum_{i\in I} \lambda_i Q_i \mid \lambda_i \in {\mathbb R}_{\geq 0} \  \forall i \in I \right\}\subset {\mathbb{R}^n}
\end{equation}
inside the FI parameter space. The cones associated to sets $I$ of cardinality $|I|=n-1$ such that the $Q_i$, $i\in I$ are linearly independent (at least classically) describe the boundaries between different phases, see e.g.~\cite{Clingempeel:2018iub}. Along each such phase boundary the unbroken gauge group, which is the stabilizer of all the $X_i$, $i\in I$ contains a single $U(1)$. 

This picture emerges from a semi-classical analysis, in which only the FI-parameters $r^a$ enter, while the $\theta$-angles remain unconstrained by the potential. Quantum effects lead to a refinement. 

On the one hand, the FI-parameters $r^a$ are subject to renormalization unless the Calabi-Yau condition
\begin{equation}
\sum_{i=1}^k Q_{ai}=0
\end{equation}
is satisfied \cite{Witten:1993yc}. In this `non-anomalous' or Calabi-Yau case the FI parameters do not run and remain  adjustable parameters of the theory.  On the other hand, quantum corrections introduce a $\theta$-dependence into the scalar potential. It turns out that the Coulomb branch only emerges at very specific values of $\theta$ on the classical phase boundaries defined by the $\text{Cone}_I$. This leads to codimension-two-loci of singular points in the K\"ahler moduli space (the space parametrized by the complexified FI parameters $r_a+i\theta_a$).
Indeed, on any such classical phase boundary, the singular loci are given by 
\begin{equation}\label{eq:sing}
\theta_I \in\{ 2\pi {\mathbb Z} + \pi S_I\}\,,\quad\text{with}\quad S_I= \sum_{Q_{I\, i}>0} Q_{I\, i} \ ,
\end{equation}
where $\theta_I$ is the $\theta$-parameter in the direction of the $U(1)$-subgroup unbroken on the phase boundary associated to $\text{Cone}_I$, and $Q_{I\, i}$ is the respective $U(1)$-charge of the chiral field $X_i$. 

Any path that avoids these singularities can be chosen to connect adjacent phases. In this way, one obtains homotopically differents paths that are specified by connected components of the space $\mathbb{R}\backslash \{ 2\pi {\mathbb Z} + \pi S_I\}$.

Our aim is to construct defects describing the transitions between the different phases of abelian GLSMs depending on the homotopy class of paths between them.

First of all, these defects have to preserve B-type supersymmetry. Namely, the perturbations captured by the FI-parameters are of K\"ahler type, i.e.~they are triggered by fields in the $(a,c)$- and $(c,a)$-rings of the theory, which preserve B-type supersymmetry on boundaries, respectively defects, while generically breaking A-type supersymmetry \cite{Brunner:2007ur}. Fusion among B-type defects (or of B-type defects with B-type boundary conditions) is non-singular and protected by supersymmetry.

To single out the sector protected by B-type supersymmetry, we will work in the limit in which the vector multiplet fields decouple. In this way, we restrict  to a topological subsector, in which the gauge symmetry persists only as an equivariance condition in the matter sector. 
This allows us to work in the setting of topological field theory, where merging of defects is well defined.
Indeed, in this setting B-type defects form a 2-category, in which their fusion is encoded as 1-composition. The relevant categories are categories of matrix factorizations, which we review in the next subsections.

Let us comment on the anomalous or non-Calabi-Yau case. In these models the FI-parameter run under a non-trivial RG flow which drives the system to an IR phase. Nonetheless, it is possible to embed a UV phase by fine-tuning of parameters \cite{Clingempeel:2018iub}. Since our defect construction is entirely in the sector protected by B-type supersymmetry, it applies to the anomalous case as well. We will illustrate this with examples in Section~\ref{sec:minmods}. 

\subsection{The Category of Matrix Factorizations}

B-type defects and boundary conditions in abelian GLSMs and their Landau-Ginzburg phases can be efficiently described by equivariant matrix factorizations.\footnote{Defects between GLSMs and geometric phases can be described by a hybrid of matrix factorizations and complexes of coherent sheaves \cite{Brunner:2021ulc}.}  

A matrix factorization of a polynomial $W\in\mathbb{C}[X_1,\dots,X_N]\coloneqq R$ is a $\mathbb{Z}_2$-graded $R$-module $P=P_0\oplus P_1$ together with an odd (i.e. of $\mathbb{Z}_2$-degree 1) endomorphsim $Q$ such that $Q^2=W \text{id}_P$.
Since $Q$ is odd it may  be written as
\begin{equation}
Q
=
\left(
\begin{matrix}
0 & p_1\\
p_0 & 0 \\
\end{matrix}
\right)
.
\end{equation}
This data can also be expressed by a two-periodic twisted chain complex
\begin{equation}
\dots\stackrel{p_0}{\longrightarrow} P_1\stackrel{p_1}{\longrightarrow} P_0 \stackrel{p_0}{\longrightarrow} P_1 \stackrel{p_1}{\longrightarrow} P_0 \stackrel{p_0}{\longrightarrow}\dots ,
\end{equation}
for which we also use the short hand notation 
\begin{equation}
\begin{tikzcd}
P: P_1 \arrow[r, "p_1", shift left] & P_0 \arrow[l, "p_0", shift left]
\end{tikzcd}.
\end{equation}
Such matrix factorizations form a category $\text{MF}(R,W)$, where morphisms between two matrix factorizations are given by the respective chain maps. The homotopy category of 
$\text{MF}(R,W)$ is denoted by $\text{HMF}(R,W)$. It describes the B-type boundary conditions in a Landau-Ginzburg model with chiral fields $X_i$ and superpotential $W$ \cite{Kapustin:2002bi,Brunner:2003dc}.

Assume that a group $G$ acts on the ring $R$ by means of ring homomorphisms such that $W$ is invariant under the action. Then, a $G$-equivariant matrix factorization of $W$ is a 
matrix factorization 
\begin{equation}
\begin{tikzcd}
P: P_1 \arrow[r, "p_1", shift left] & P_0 \arrow[l, "p_0", shift left]
\end{tikzcd},
\end{equation}
of $W$ together with representations $\rho_{P_i}:G\rightarrow \text{GL}(P_i)$ of $G$ on $P_i$ compatible with the representation $\rho$  on the ring $R$, i.e.
\begin{equation}
\rho_{P_i}(g)(rp)=\rho(g)(r)\rho_{P_i}(g)(p),\ \forall\ g\in G,\ r\in R,\ p\in P_i
\end{equation}
such that
\begin{equation}
\rho_{P_0}(g)p_1\rho_{P_1}^{-1}(g)=p_1,\ \rho_{P_1}(g)p_0\rho_{P_0}^{-1}(g)=p_0,\ \forall\ g\in G.
\end{equation}
The representations $\rho_i$ induce representations on chain maps between $G$-equivariant matrix factorizations. $G$-invariant chain maps are the homomorphisms between $G$-equivariant matrix factorizations. We denote the respective category of matrix factorizations by $MF(R,W)^G$, and the respective homotopy category by $HMF(R,W)^G$. 

In the case $G=U(1)^n$ the representations $\rho_i$ can be specified by means of the weights (charges) of the respective representations. Indeed, $U(1)^n$-equivariant matrix factorizations can be regarded as $\mathbb{Z}^n$-graded matrix factorizations. 

Later in the paper we will make extensive use of the connection between matrix factorizations and maximal Cohen-Macaulay modules \cite{eisenbud}. Instead of working with a matrix factorization
\begin{equation}
\dots\stackrel{p_0}{\longrightarrow} P_1\stackrel{p_1}{\longrightarrow} P_0 \stackrel{p_0}{\longrightarrow} P_1 \stackrel{p_1}{\longrightarrow} P_0 \stackrel{p_0}{\longrightarrow}\dots ,
\end{equation}
of $W\in R$, we will consider the associated 
$S:=R/WR$-module $\mathcal{P}\coloneqq \text{coker}(p_1:P_1\otimes_R S\rightarrow P_0\otimes_R S)$, which is maximal Cohen-Macaulay. 

Indeed, any maximal Cohen-Macualay module over $S$ admits a minimal free resolution which is two-periodic and defines a matrix factorization of $W$. More generally, any Cohen-Macaulay module
admits a free resolution which turns two-periodic after finitely many steps. The two-periodic part then gives rise to a matrix factorization.

This correspondence between matrix factorizations and Cohen-Macaulay modules can be extended to the equivariant, respectively graded setting. In particular, 
$\mathbb{Z}^n$-graded matrix factorizations are related to $\mathbb{Z}^n$-graded Cohen-Macaulay modules.

\subsection{Defects in GLSMs and Orbifold Phases}\label{sec:ID}

A B-type supersymmetric defect separating two GLSMs

\begin{equation}
(G_1, \mathbb{C}^{k^{(1)}},(r^{(1)},\theta^{(1)}),W^{(1)})
\quad\text{and}\quad
(G_2,\mathbb{C}^{k^{(2)}},(r^{(2)},\theta^{(2)}),W^{(2)}),
\end{equation}
can be represented by a $G_1\times G_2$-equivariant matrix factorization of the difference $W^{(1)}-W^{(2)}$ of the respective superpotentials, or more precisely by an object 
in $HMF(W^{(1)}-W^{(2)},R)^{G_1\times G_2}$ where $R=\mathbb{C}[X_1^{(1)},\ldots,X_{k^{(1)}}^{(1)},X_1^{(2)},\ldots,X_{k^{(2)}}^{(2)}]$ is the ring of chiral fields of the two models.\footnote{Note that a boundary condition (D-brane) is just a defect separating a GLSM from the trivial GLSM (no chiral fields and zero superpotential, trivial gauge group).} 

In fact, defects are not just objects in a category but also 1-morphisms in a 2-category. In particular they can be composed, which corresponds to the fusion of line defects. In our setting, this fusion is realized by the tensor product of matrix factorizations over the ring of bulk fields of the model between the two defects. In more detail, the tensor product of two matrix factorizations
$P$ of $W\in R$ and $P'$ of $W'\in R'$ is just given by the tensor product of the respective two-periodic twisted complexes:
\begin{equation}
P\otimes_\mathbb{C} P'\coloneqq \underbrace{(P_0\otimes_\mathbb{C}P'_0)\oplus (P_1\otimes_\mathbb{C}P'_1)}_{(P\otimes P')_0}\oplus \underbrace{(P_1\otimes_\mathbb{C}P'_0)\oplus (P_0\otimes_\mathbb{C}P'_1)}_{(P\otimes P')_1},
\end{equation}
with the twisted differential
\begin{equation}
Q\coloneqq
\left(
\begin{matrix}
0 & 0 & p_1 & -p'_1\\
0 & 0 & p'_0 & p_0\\
p_0 & p'_1 & 0 & 0\\
-p'_0 & p_1 & 0 & 0\\
\end{matrix}
\right)
\end{equation}
Since the twists add under the tensor product, it is a matrix factorization of the sum $W+W'$. 

Now, let us consider three models with chiral fields $X_i^{(a)}$ and superpotential $W_a(X_i^{(a)})$, $a=1,2,3$ (we consider trivial gauge group for the moment) and a defect $P^{(1)}$ between model $2$ and model $1$ and $P^{(2)}$ between model $3$ and model $2$ as depicted in \eqref{pic:twodefects} below.
Then $P^{(1)}$ is represented by a matrix factorization of $W_1(X_i^{(1)})-W_2(X_i^{(2)})$ over $R^{(1,2)}$ and 
$P^{(2)}$ by a matrix factorization of $W_2(X_i^{(2)})-W_3(X_i^{(3)})$ over $R^{(2,3)}$. Here $R^{(a,b)}:=R^{(a)}\otimes_\mathbb{C} R^{(b)}$ with $R^{(a)}=\mathbb{C}[X_i^{(a)}]$ denote the ring of chiral fields of the bulk models adjoining the respective defects.

\begin{equation}\label{pic:twodefects}
\tikz[baseline=20,scale=0.7]{
		\fill[WScolor light] (0,0) rectangle (3,1.5);
		\fill[WScolor] (3,0) rectangle (6,1.5);
		\fill[WScolor light] (6,0) rectangle (9,1.5);
		\draw[defect] (3,0) -- (3,1.5);
		\draw[defect] (6,0) -- (6,1.5);
		\node at (4.5,.75) {$W_2(X_i^{(2)})$};
		\node at (1.5,.75) {$W_3(X_i^{(3)})$};
		\node at (7.5,.75) {$W_1(X_i^{(1)})$};
		\node at (3.1,-.5) {$P^{(2)}$};
		\node at (6.1,-.5) {$P^{(1)}$}
	}
\end{equation}

The fusion of $P^{(1)}$ and $P^{(2)}$ over the model $2$ squeezed in between them is then given by the tensor product 
\begin{equation}
P^{(1)}\otimes_{R^{(2)}} P^{(2)}
\end{equation}
over the field of bulk fields of model $2$, regarded as matrix factorization over the ring
$R^{(1,3)}$. Note that this matrix factorization still involves the chiral bulk fields $X_i^{(2)}$ of the model squeezed in between the defects. Hence, a priori it is a matrix factorization of infinite rank. It can however be shown to be isomorphic to a matrix factorization of finite rank. 
Physically, the bulk degrees of freedom of the squeezed in model are 
promoted to defect degrees of freedom. For more information about defect fusion in this setup we refer to \cite{Brunner:2007qu}.

This formula is modified only slightly in the presence of gauge groups. Suppose that 
the models $a$ have gauge group $G_a$ for $a=1,2,3$.  Then the matrix factorizations $P^{(1)}$ and $P^{(2)}$ representing defects between the models are equivariant with respect to $G_1\times G_2$, respectively $G_2\times G_3$. The tensor product 
$P^{(1)}\otimes_{R^{(2)}} P^{(2)}$ then carries a representation of $G_2$, and the fusion of the respective defects is given by the $G_2$-invariant sub matrix factorization
\begin{equation}
P^{(1)}\ast P^{(2)}=\left[P^{(1)}\otimes_{R^{(2)}} P^{(2)}\right]^{G_2}
\end{equation}
More details on fusion in the equivariant setup can be found in \cite{Brunner:2007ur}.

On the level of Cohen-Macaulay modules, fusion is represented in a similar fashion, by taking the invariant part with respect of the intermediate gauge group of the tensor product of Cohen-Macaulay modules over the intermediate ring. 

The starting point of our construction is a very specific defect, namely the identity defect of the GLSM.
The identity defect (sometimes also referred to as invisible defect) is a defect that separates a given theory from itself in a trivial way in the sense that its presence does not affect correlation functions. In particular fusion with this defect is trivial. 
The identity defect enforces trivial gluing conditions $X_i^{(1)}=X_i^{(2)}$ on the fields on either side. In the 
case of trivial gauge group  $G=G_1=G_2$, it is represented by a matrix factorization of Koszul-type with factors $(X_i^{(1)}-X_i^{(2)})$ -- it is a tensor product  of rank-1 matrix factorizations with $p_1=(X_i^{(1)}-X_i^{(2)})$ for each chiral bulk field $X_i$. One can associate the Cohen-Macaulay module\footnote{Note that if the model has more than one chiral field, this module is not maximal Cohen-Macaulay -- the resolution is not two-periodic from the start, but only after the number of chiral fields minus one steps.} 
\begin{equation}\label{eq:id-module}
\mathcal{I}=\frac{S^{(1,2)}}{\langle (X_i^{(1)}-X_i^{(2)})_{i=1,\dots,k}\rangle}\,,
\end{equation}
where
\begin{equation}
S^{(1,2)}=R^{(1,2)}/(W(X_i^{(1)})-W(X_i^{(2)}))R^{(1,2)}.
\end{equation}
If the model has a non-trivial gauge group $G$, the matrix factorization for the identity defect has to be modified. Namely, the factors $(X_i^{(1)}-X_i^{(2)})$ are not equivariant with respect to the action of the product $G\times G$ of the gauge groups on the left, respectively right side of the defect. To make them equivariant, one has to introduce intertwiners of the representations, which is accomplished by tensoring the Koszul factorization by the regular representation of the gauge group. For $G=U(1)$, the regular representation is given by
\begin{equation}
V_\text{reg}^{U(1)}=\frac{\mathbb{C}[\alpha,\alpha^{-1}]}{\langle \alpha\alpha^{-1}-1 \rangle}\,.
\end{equation}
Here $\alpha$ corresponds to a defect field which has charges $1$ and $-1$ under the $U(1)$ gauge groups on the left, respectively right of the defect. The inverse field has the inverse charges.

For $G=U(1)^n$, one such field $\alpha_i$ has to be introduced for every $U(1)$-factor, i.e. 
\begin{equation}
V_\text{reg}^{U(1)^n}=\frac{\mathbb{C}[\alpha_1,\alpha^{-1}_1,\ldots,\alpha_n,\alpha^{-1}_n]}{\langle (\alpha_a\alpha^{-1}_a-1)_{a=1,\ldots,n}\rangle}\, .
\end{equation}
$\alpha_a$ has charges $1$ and $-1$ under the $a$th $U(1)$ and $0$ under all the other $U(1)$s. The inverse $\alpha_a^{-1}$ has the opposite charges. These fields can now be used to render the Koszul factors in \eqref{eq:id-module} equivariant
\begin{equation}
(\alpha_1^{-Q_{1\,i}}\dots\alpha_n^{-Q_{n\,i}}X^{(1)}_i-X^{(2)}_i)\,.
\end{equation}
Here $Q_{a\,i}$ is the charge matrix specifying the gauge representations on the chiral matter fields.  

The Cohen-Macaulay module associated to the identity defect in the GLSM
$(U(1)^n,\mathbb{C}^k,(r,\theta),W)$ is then given by 
\begin{equation} \label{eq:GLSM-id}
\mathcal{I}_{\text{GLSM}}\coloneqq \frac{S^{(1,2)}\otimes V^{U(1)^n}_\text{reg}}
{\langle (\alpha_1^{-Q_{1\,i}}\dots\alpha_n^{-Q_{n\,i}}X_i^{(1)}-X_i^{(2)})_{i=1,\dots,k},\rangle}
\end{equation}

The construction of the identity defect in Landau-Ginzburg orbifold models is completely analogous. One only has to replace the regular representation of the gauge group $G=U(1)^n$ in \eqref{eq:GLSM-id} by the regular representation of the respective finite orbifold group. For instance, in the important case of cyclic orbifold groups $\mathbb{Z}_d$, the regular representation is given by 
\begin{equation}
V_\text{reg}^{{\mathbb Z}_d}= \frac{\mathbb{C}[\alpha]}{\langle \alpha^d-1 \rangle}\, .
\end{equation}

For more details on the representation of the identity defect by means of matrix factorizations and Cohen-Macaulay modules see \cite{Brunner:2007qu} for the non-equivariant case, \cite{Brunner:2007ur} for the case of finite gauge groups and \cite{Brunner:2021cga} for the case of gauge group $U(1)$.

\subsection{Orbifold Lift}

In this section, we consider GLSMs that admit Landau-Ginzburg phases. In the latter, the D-term equations require that some of the chiral fields of the GLSM obtain a non-trivial vacuum expectation value, which in turn breaks the gauge symmetry to a finite subgroup $H\subset G$ of the gauge group $G=U(1)^n$. Let us denote the fields in such a way that $X_{N+1} \dots X_k$ aquire a non-trivial vev in the LG phase. Then $H$ is the stabilizer of these fields, and the Landau-Ginzburg model has chiral fields given by $X_1,\ldots,X_N$ and superpotential $W_\text{LG}$ obtained by inserting the vacuum-expectation values for the fields $X_{N+1} \dots X_k$ into the superpotential $W$ of the GLSM.

Our aim is to construct the defects  $T^i_\gamma$ ($i$ labelling the LG phase and $\gamma$ the homotopy class of a path in the GLSM parameter space) which lift the 
Landau-Ginzburg phase to the GLSM, c.f. Section~\ref{sec:outline}.
 Such defects are B-type defects mediating between the Landau-Ginzburg model and the GLSM, and can therefore be represented by objects in 
$HMF(W-W_\text{LG})^{G\times H}$.

Before delving into the construction of the $T^i_\gamma$, let us remark on the relationship between the identity defect of the GLSM and the one in the phase. Indeed, setting all the fields $X_i$, $N<i\leq k$ to their vevs (which we choose to be $1$) on both sides of the GLSM identity defect yields the identity defect of the Landau-Ginzburg phase. Namely, setting $X_i^{(1)}= X_i^{(2)}=1$ for $N<i\leq k$ changes the respective Koszul factors to 
\begin{equation}\label{eq:gesetzt}
\alpha_1^{-Q_{1\,i}}\dots\alpha_n^{-Q_{n\,i}}=1, \quad  i \in \{ N+1, \dots k \} \ .
	\end{equation}
Dividing out $V_\text{ref}^{U(1)^n}$ by this relation precisely yields the regular representation $V_\text{reg}^H$ of $H$. Thus quotienting by $X_i^{(1)}= X_i^{(2)}=1$ for $N<i\leq k$ in ${\mathcal I}_\text{GLSM}$ produces the module associated to the identity defect in the Landau-Ginzburg phase. 

Conversely, starting with the identity defect of the LG phase, one may ask how to lift to a defect of the GLSM. This can be done by re-introducing the variables $X_i$, $N<i\leq k$ on both sides of the defect and lifting the respective representations of $H$ to representations of $G$. Indeed, there are many different such lifts which we refer to as ``lifted identities". None of these lifted identities is the GLSM identity defect. 

We now turn to the construction of defects $T_\gamma^{i}$ and $R_\gamma^\text{i}$ for Landau-Ginzburg phases.
These defects have to satisfy
\begin{equation} \label{eq:LGcond}
R_\gamma^i \otimes T_\gamma^i= \id_i
\end{equation}
We will omit the superscript $i$ in the following, or replace it by $\text{LG}$ to emphasize that the phase is an LG phase. 

Our starting point is the identity defect of the GLSM \eqref{eq:GLSM-id}. Motivated by the observation that setting the fields to their expectation values on both sides yields the identity of the phase, we factorize the identity defect of the phase over the GLSM:
\begin{equation}
R_\infty \otimes T_\infty = \id_\text{LG} \ ,
\end{equation}
where the fusion product is the $U(1)^n$ equivariant tensor product over the GLSM fields squeezed in in the middle.
The defects $R_\infty$ and $T_\infty$ are obtained by setting the fields $X_i$, $N<i\leq k$ to their vevs, but only on one side of the GLSM identity defect. For example, to obtain $T_\infty$ we set 
\begin{equation}\label{eq:settoone}
X_i^{(2)}= 1 \,,\; N<i\leq k
\end{equation}
in the GLSM identity defect and arrive at the setting depicted in \eqref{pic:pushdown}. 
\begin{equation}\label{pic:pushdown}
\tikz[baseline=10]{
	\fill[WScolor] (-2,0) rectangle (4,1);
	\draw[densely dashed] (1,0) -- (1,1);
	\node at (1.55,.2) {$I_\text{GLSM}$};
	\node at (-.5,.5) {$W(X^{(1)}_i)$};
	\node at (2.5,.5) {$W(X^{(2)}_i)$};
}\;\tikz[baseline=-4]{
\node at (0.7,0.4) {\eqref{eq:settoone}};
	\path (0,0) edge[snake it] (1.5,0);
}\;\tikz[baseline=10]{
	\fill[WScolor] (-2,0) rectangle (1,1);
	\fill[WScolor light] (1,0) rectangle (4,1);
	\draw[defect] (1,0) -- (1,1);
	\node at (-.5,.5) {$W(X_i^{(1)})$};
	\node at (2.5,.5) {$W(X_i^{(2)})$};
	\node at (1.3,.2) {$T_\infty$};
}
\end{equation}
On the level of modules, imposing \eqref{eq:settoone} in \eqref{eq:GLSM-id}
yields a new module 
 $\mathcal{T}_\infty$ with relations 
\begin{eqnarray}
		\alpha_1^{-Q_{1\,i}}\dots\alpha_n^{-Q_{n\,i}}X_i^{(1)} = X_i^{(2)},\ \ &&1\leq i\leq N\nonumber
\\
	\alpha_1^{-Q_{1\,i}}\dots\alpha_n^{-Q_{n\,i}}{X_i^{(1)}} = 1,  \ \ &&N<i\leq k\label{eq:1rel}
\end{eqnarray}
Thus
\begin{equation}
{\mathcal{T}}_\infty = \frac{S^{(1,2)}\otimes V_\text{reg}}
{\langle (\alpha_1^{-Q_{1\,i}}\dots\alpha_n^{-Q_{n\,i}}X_i^{(1)}-X_i^{(2)})_{i=1,\dots,N}\rangle
\langle (\alpha_1^{-Q_{1\,i}}\dots\alpha_n^{-Q_{n\,i}}{X_i^{(1)}}-1)_{i=N+1, \dots, k)} \rangle}
\end{equation}
Similarly, we obtain a module $\mathcal{R}_\infty$ by setting fields on the left hand side of the defect to their expectation value. 

The resulting modules $\mathcal{T}_\infty$ however do not have the right properties for the lift defects discussed in Section~\ref{sec:outline} above. For one thing, $\mathcal{T}_\infty$ is not finitely generated and hence cannot be obtained by lifting the identity defect of the Landau-Ginzburg phase on the left side to the GLSM. What is more, 
${\mathcal T}_\infty$ is unique and hence does not depend on a homotopy class of paths in moduli space.
For the case of $U(1)$ gauge groups, this problem was solved in \cite{Brunner:2021cga}. There it was shown that the lift defects are obtained from ${\mathcal T}_\infty$ by a cutoff procedure which renders the module finitely generated. The choice involved in the cutoff procedure exactly corresponds to the choice of homotopy class of paths in parameter space. 
In the following, we will generalize this to higher rank abelian gauge groups and give a concrete construction of the desired transition defects. 

Given a path $\gamma$ in parameter space. This path crosses a number of phase boundaries associated to cones Cone$_{I_s}$, $s=1,\ldots,m$ as defined in \eqref{eq:cone}. On each one, a $U(1)$-subgroup $U(1)_{I_s}$ is preserved. To construct the lift defects, we now impose 
cutoffs in $\mathcal{T}_\infty$
\begin{equation}\label{eq:chargecutoff}
Q^L_{I_s}\leq N_{I_s}
\end{equation}
of the respective charges $Q^L_{I_s}$ of the gauge group on the left of the defect for each such transition. More precisely, one considers the submodule $\mathcal{T}_{N_{I_1},\ldots,N_{I_m}}\subset {\mathcal T}_\infty$ generated by all the generators whose charges satisfy \eqref{eq:chargecutoff}. 
We claim that the associated defects ${T}_{N_{I_1},\ldots,N_{I_m}}$ are the respective lift defects. 
While this does not offer an apriori assignment of defects to homotopy classes of paths, we observe that 
the choice of cutoff parameters  $N_{I_1},\ldots,N_{I_m}$ are in one-to-one correspondence with the homotopy classes of paths with the chosen phase transitions. 
(For each transition, 
the different paths must avoid the singular locus \eqref{eq:sing}, and a homotopy class is specified by a connected component of $\mathbb{R}\backslash(2\pi\mathbb{Z}+\pi S_I)$.) 
Thus, our construction provides a lift defect for every homotopy class of paths not encircling the singular points on the phase boundaries. 

Indeed, the lift defects satisfy $R_\infty \otimes T_{N_{I_1},\ldots,N_{I_m}}= \id_\text{LG}$ and hence are  ``lifted identities'' from the perspective of the orbifold phase. Moreover, when the path crosses $r$ phase boundaries ($r$ is the rank of the gauge group), the respective cutoffs render the module finitely generated. Indeed, due to the relations \eqref{eq:1rel} in ${\mathcal T}_\infty$, any cutoff \eqref{eq:chargecutoff} automatically leads to a lower limit on the respective charges of necessary generators of ${\mathcal T}_{N_{I_1},\ldots,N_{I_m}}$:
\begin{equation}\label{eq:chargeband}
N_{I_s}-M_{I_s}<Q^L_{I_s}\leq N_{I_s}.
\end{equation}
Here $M_{I_s}$ is an integer associated to $I_s$. This will be explained in more detail in the examples in Sections~\ref{sec:examples} and \ref{sec:minmods}.
As will be outlined in Section~\ref{sec:grr} below, this exactly reproduces the band restriction rule for D-brane transport as put forward in \cite{herbst2008}.

From the lift defects $T_{N_{I_1},\ldots,N_{I_m}}$ constructed in this way, one can then obtain the defects describing the transition between the LG phase and any other phase crossed by the path $\gamma$. This is done by pushing the GLSM to the respective phase on the left of the defect, which is the same as fusion with the respective defect $R_\infty$. If the target phase is another Landau-Ginzburg phase, this just involves setting fields to their vevs. For geometric phases, this is somewhat more complicated. It involves expanding the lift defect $T_{N_{I_1},\ldots,N_{I_m}}$ into a complex of matrix factorizations and interpreting it as a hybrid between a matrix factorization and a complex of coherent sheaves on the target space. These steps were performed for one parameter models in \cite{Brunner:2021ulc}. We omit them here, focussing on the construction of the lift defects. We will construct them for concrete examples in 
Sections~\ref{sec:examples} and \ref{sec:minmods} below.

\subsection{Comparison with the Grade Restriction Rule}\label{sec:grr}

To put our construction into perspective, we would like to compare it to results on D-brane transport on K\"ahler moduli space in \cite{herbst2008}. Indeed, 
fusion of transition defects with D-branes (boundary conditions) describes the behavior of the latter under the respective phase transitions. In this way, the defects constructed above can be used to describe D-brane transport between different phases in K\"ahler moduli space. Fusion of D-branes in the LG phase with the defects $T_{N_{I_1},\ldots,N_{I_m}}$ lifts these D-branes to the GLSM in a way compatible with transport along a path associated to the choice of cutoff parameters. It turns out that this matches precisely with the band restriction rule proposed in \cite{herbst2008} for the Calabi-Yau case. 

The latter states the following: A path between two adjacent phases has to avoid the singular locus \eqref{eq:sing} in K\"ahler moduli space, i.e. it crosses the phase boundary at  $\theta_I\in \mathbb{R}\backslash\{2\pi\mathbb{Z}+\pi S_I\}$. The choice of any such $\theta_I$  gives rise to a window
\begin{equation}\label{eq:GRRHHP}
\mathbb{Z}\cap\left\{-\frac{\theta_I}{2\pi}+\left(-\frac{S}{2},\frac{S}{2}\right)\right\}
\end{equation}
of consecutive integers, which
only depends on the connected component of possible $\theta_I$ and hence on the homotopy class of paths from one phase to the adjacent one. According to the band restriction rule (see Section 7.3.2.~of \cite{herbst2008}), D-branes built from Wilson line branes ${\mathcal W}(q)$ can be transported straightforwardly along a chosen path in K\"ahler moduli if and only if their charges $q$ under the $U(1)$s unbroken at all phase boundaries traversed lie in the respective windows.

The matrix factorizations associated to the lift defects $T_{N_{I_1},\ldots,N_{I_m}}$ have the property that the generators of their underlying modules have charges (under the gauge group on the left of the defect) lying in the band \eqref{eq:chargeband}. This means that fusing any boundary condition (D-brane) in the LG phase with 
${T}_{N_{I_1},\ldots,N_{I_m}}$ produces only GLSM branes whose charges lie in this charge band.
Thus, lifting LG-branes into the GLSM with ${T}_{N_{I_1},\ldots,N_{I_m}}$ produces GLSM branes in that charge band. As we will see concretely in the examples discussed in Section~\ref{sec:examples}, the charge bands singled out by the defect construction precisely match those of the band restriction rule in \cite{herbst2008}. Note that while our construction requires the introduction of the upper bounds on the charges, the lower bounds automatically follow from it. In particular, the size of the bands is completely determined from the construction.

Note that a path from the small volume phase ($r_i<<0$ for all $i$)  to the large volume phase ($r_i>>0$ for all $i$) crosses at least $r$ phase boundaries, where $r$ is the rank of the gauge group. Hence, the corresponding band restriction rule restricts to a finite set of possible charges. In the language of defects and corresponding modules, this matches the fact that ${\mathcal T}_{\infty}$ gets truncated to a finite submodule in this case.

For the anomalous case, a generalization of the grade restriction rule has been discussed in \cite{Hori:2013ika,Clingempeel:2018iub}. Starting from the UV phase, D-branes are lifted into a `large window', and D-branes with charges in a `small window' that is a subset of the large window survive the flow to an IR phase. The location of the small window inside the large window can be shifted by symmetry and monodromy. Our defect construction applies to the anomalous case as well, the lift defects lift D-branes into the large window, and the small window arises automatically when pushing to the IR phase on the other side of the defect. We will confirm this for a class of examples, the Landau-Ginzburg models with superpotential $W=X^d$ in Section~\ref{sec:minmods}. 
\section{Non-Anomalous Examples}\label{sec:examples}

In this section we will apply the construction of lift defects described in  the previous section to two concrete examples of non-anomalous GLSMs with higher-rank abelian gauge groups. For simplicity we will choose examples with zero superpotential. The relevant defects are described by matrix factorizations of $0$, which correspond to honest (as opposed to twisted) complexes, and hence there is a direct connection with modules over polynomial rings. 

The first example is the $A_{N-1}$-model. It is a GLSM with $U(1)^{N-1}$ gauge group, whose orbifold phase is associated to the orbifold ${\mathbb C}^{2}/\mathbb{Z}_N$. Here $\mathbb{Z}_N$ acts by opposite phase multiplication on the two coordinates of ${\mathbb C}^{2}$ such that the Calabi-Yau condition is satisfied. The geometric phase of this model is the sigma model on the resolution of the corresponding $A_{N-1}$-singularity by a chain of ${N-1}$ $\mathbb{P}^1$'s, whose volumes are determined by the exactly marginal FI parameters. 

The second example is a $U(1)^2$ GLSM which has four phases, one of which is a $\mathbb{C}^5/\mathbb{Z}_8$-orbifold phase and one which is a non-linear sigma model on the total space of the line bundle ${\mathcal O}(-8)$ over the weighted projective space ${\mathbb P}_{(11222)}$.

These models are well studied. In particular, the D-brane transport between different phases in these models has been investigated in \cite{herbst2008} (where the two models are referred to as example (D) and example (C), respectively). Now, D-brane transport is easy to describe in our framework, since it is just given by fusion of the D-branes with the transition defects between the respective phases. So after constructing the lift defects of the orbifold phases in these examples, we will show that fusion of D-branes with these defects reproduces the results on D-brane transport from \cite{herbst2008}. 
The latter are formulated in terms of the grade restriction (for rank $1$ gauge theories) or band restriction (higher rank) rule. 

This section is organized as follows: We will start by studying two different two-parameter models, the $A_2$-model and the model with orbifold phase $\mathbb{C}^5/\mathbb{Z}_8$. In both cases we compare our results with the band restriction rule and find agreement. 
Subsequently, we will generalize the discussion of the $A_2$-model to the $N-1$-parameter $A_{N-1}$-model  for arbitrary $N$.

Throughout this section, we will distinguish variables on the left and right of the defect by adding a prime, while the respective gauge groups are distinguished by a superscript $L/R$.

\subsection{The $A_2$ Model}\label{sec:a2}

We start out by studying lifts of the orbifold phase of the $A_2$-GLSM. The latter is specified by 
the data
\begin{equation}
	\text{GLSM}_{A_2}\coloneqq(U(1)_1\times U(1)_2,\ V,\ (r,\theta),\ W=0),
\end{equation}
where the representation $V$, can be read off from the charge assignment of the chiral matter fields specified in Table~\ref{A_2Matter} below.

\begin{table}[H]
\centering
	\begin{tabular}{c|cccc}
	& $X_1$ & $X_2$ & $X_3$ & $X_4$\\
	\hline
	$Q_1$ & 1 & -2 & 1 & 0\\
	$Q_2$ & 0 & 1 & -2 & 1\\
	\end{tabular}
\caption{Matter content of the $A_2$-model. $Q_j$ denotes the respective $U(1)_j$-charge.}
\label{A_2Matter}
\end{table}
\noindent

This model exhibits an orbifold phase for $r_1,\ r_2\rightarrow -\infty$.
In this phase the fields $X_2$ and $X_3$ acquire a vev and the low energy theory is the orbifold theory $\mathbb{C}^2/\mathbb{Z}_3$ with fields $X_1$ and $X_4$ whose $\mathbb{Z}_3$-charges are $1$ and $-1$ respectively. The model also has a large volume phase, described by a sigma model on the resolution of the $\mathbb{C}^2/\mathbb{Z}_3$ singularity, where both of the 2-spheres in the exceptional divisor are blown up. Apart from these there are two mixed phases related to partial resolutions of the singularity, where only one of the two 2-spheres is blown up. 

The phase diagram of $\text{GLSM}_{A_2}$ is depicted in Figure~\ref{A_2Phases} below.
Here $(11)$ denotes the--unresolved--orbifold phase and $(00)$ denotes the large volume phase, where both 2-spheres in the exceptional divisor are blown up. 
The intermediate phases in which only one of the two 2-spheres is blown up are denoted by $(01)$ and $(10)$, respectively.

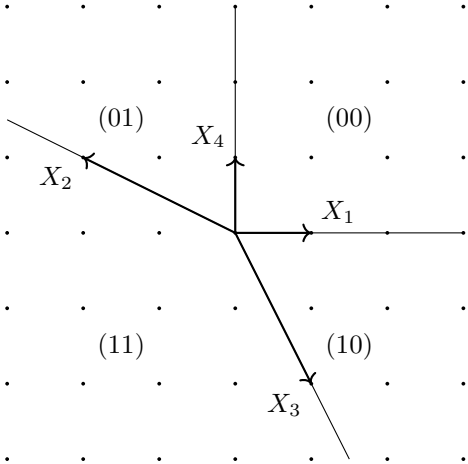
\begin{figure}[h]
\center
\begin{tikzpicture}
\foreach \x in {-3,-2,...,3} {
\foreach \y in {-3,-2,...,3} {
\fill[color=black] (\x,\y) circle (0.25mm);
}
}
\draw[thick,->] (0,0) -- (1,0) node[anchor=south west] {$X_1$};
\draw[thick,->] (0,0) -- (-2,1) node[anchor=north east] {$X_2$};
\draw[thick,->] (0,0) -- (1,-2) node[anchor=north east] {$X_3$};
\draw[thick,->] (0,0) -- (0,1) node[anchor=south east] {$X_4$};
\draw[thick] (1.5,1.5) node[] {$(00)$};
\draw[thick] (-1.5,-1.5) node[] {$(11)$};
\draw[thick] (1.5,-1.5) node[] {$(10)$};
\draw[thick] (-1.5,1.5) node[] {$(01)$};
\draw[thin,-] (0,0) -- (0,3);
\draw[thin,-] (0,0) -- (3,0) ;
\draw[thin,-] (0,0) -- (-3,1.5);
\draw[thin,-] (0,0) -- (1.5,-3);
\end{tikzpicture}
\caption{Phase diagram of $\text{GLSM}_{A_2}$. Phases are denoted by $(i_1i_2)$ where $i_j=0$ if the $j$th exceptional 2-sphere is blown up, and $i_j=1$ if it is not blown up. The phase boundaries denoted by $X_i$ are located at $\text{Cone}_{\{i\}}$.}
\label{A_2Phases}
\end{figure}
\noindent
The phase boundary $(11)\leftrightarrow (01)$ is located at 
\begin{equation} 
\text{Cone}_{\{2\}}=\left\{
r=\lambda
	\begin{pmatrix}
	-2\\
	1
	\end{pmatrix}\Big|\  
\lambda\in\mathbb{R}_{>0}\right\}.
\end{equation}
The D-term equation forces $X_2$ to acquire a vev on this phase boundary, and the isotropy group of the latter is given by
\begin{equation}
\{(g,g^2)\vert g\in U(1)\}\cong U(1).
\end{equation}
This is the $U(1)$ unbroken on the entire phase boundary.
Analogously one obtains the unbroken gauge groups at the other phase boundaries:

\begin{center}
\begin{tabular}{ccc}
phase boundary & location & unbroken gauge group\\
\hline
$(11)\leftrightarrow (01)$&$\text{Cone}_{\{2\}}$&$\{(g,g^2)\,|\,g\in U(1)\}$\\
$(11)\leftrightarrow (10)$&$\text{Cone}_{\{3\}}$&$\{(g^2,g)\,|\,g\in U(1)\}$\\
$(00)\leftrightarrow (01)$&$\text{Cone}_{\{4\}}$&$U(1)_1$\\
$(00)\leftrightarrow (10)$&$\text{Cone}_{\{1\}}$&$U(1)_2$
\end{tabular}
\end{center}
A more detailed discussion about the $A_2$-model can be found in \cite{herbst2008}.\footnote{It is example (D) with $N=3$.}

\subsubsection{The GLSM Identity Defect}

The identity defect of GLSM$_{A_2}$ is associated to the $\mathbb{C}[X_1,\dots,X_4,X'_1,\dots,X'_4]$-module
\begin{equation}
\mathcal{I}_{\text{GLSM}_{A_2}}\coloneqq\mathbb{C}[X_1,\dots,X_4,X'_1,\dots,X'_4,\alpha_1,\alpha_1^{-1},\alpha_2,\alpha_2^{-1}]/\langle (\alpha_1^{-Q_{1\,i}}\alpha_2^{-Q_{2\,i}}X_i-X_i'),(\alpha_i\alpha^{-1}_i-1) \rangle. 
\label{A_2GLSMIdMod}
\end{equation}
Here the fields $X_i$ and $X_i^\prime$ denote the chiral fields of the model on the left, respectively the right of the defect. We also introduced auxiliary defect fields $\alpha_i$ and $\alpha^{-1}_i$, for each of the $U(1)$ gauge groups satisfying $\alpha_i\alpha_i^{-1}=1$. The $\alpha_i$ are charged under both, the gauge groups on the left and the right of the defect.
The gauge-charges of the various fields are given in Table~\ref{A_2GLSMId} below.

\begin{table}[H]
\centering
	\begin{tabular}{c|cccccccccccc}
	& $X_1$ & $X_2$ & $X_3$ & $X_4$ & $X'_1$ & $X'_2$ & $X'_3$ & $X'_4$ & $\alpha_1$ & $\alpha^{-1}_1$ & $\alpha_2$ & $\alpha^{-1}_2$\\
	\hline
	$Q^L_1$ & 1 & -2 & 1 & 0 & 0 & 0 & 0 & 0 & 1 & -1 & 0 & 0\\
	$Q^L_2$ & 0 & 1 & -2 & 1 & 0 & 0 & 0 & 0 & 0 & 0  & 1 & -1\\
	$Q^R_1$ & 0 & 0 & 0 & 0 & 1 & -2 & 1 & 0 &-1 & 1  & 0 & 0\\		
	$Q^R_2$ & 0 & 0 & 0 & 0 & 0 & 1 & -2 & 1 & 0 & 0  &-1 & 1\\
	\end{tabular}
\caption{Fields of the identity defect of the $A_2$-model. $Q^L_j$ denotes the charges of the fields under the left gauge group $U(1)^L_j$ and $Q^R_j$ denotes the charges of the fields under the right gauge group $U(1)^R_j$.}
\label{A_2GLSMId}
\end{table}
\noindent
As an aside note that pushing the GLSM identity defect to the orbifold on both sides requires setting $X_2=X_2'=X_3=X_3'=1$ in $\mathcal{I}_{\text{GLSM}_{A_2}}$. 
This imposes the relations $\alpha_1^{-2} \alpha_2=1$ and $\alpha_2^{-2}\alpha_1=1$, thereby realizing the gauge symmetry breaking to the subgroup ${\mathbb Z}_3\subset U(1)\times U(1)$ on the level of rings describing the regular representation. 
The result is the module
\begin{equation}\label{eq:orbidA2}
\mathcal{I}_{\text{orb }} = \mathbb{C}[X_1, X_4, X_1', X_4', \alpha_1]/\langle (\alpha_1^3=1), (X_1\alpha_1^{-1}-X_1'), (X_4 \alpha_1^{-2}-X_4') \rangle 
\end{equation}
associated to the identity defect in the orbifold model $\mathbb{C}^2/\mathbb{Z}_3$.
It is built on $3$ generators $e_i=\alpha_1^{i-1}$, of $\mathbb{Z}_3^L\times\mathbb{Z}_3^R$-charges\footnote{Here, $[\bullet]_d$ denotes the ${\mathbb Z}_d$-reduction of the integer written in brackets.} $([i-1]_3,[-i+1]_3)$ satisfying relations
\begin{equation}\label{eq:A2orbrel}
X_1' e_1= X_1 e_2, \ \  X_1' e_2= X_1 e_3, \ \ X_1' e_3= X_1 e_1, \quad X_4' e_1= X_4 e_3, \ \ X_4' e_2= X_4 e_1, \ \ X_4' e_3 = X_4 e_2.
\end{equation}

\subsubsection{Orbifold lift}

The defects $T$ that lift the orbifold phase to the GLSM are obtained by pushing the GLSM on the right side of the GLSM identity defect to the orbifold phase. To do this, we first set  $X'_2=X'_3=1$
in $\mathcal{I}_{\text{GLSM}_{A_2}}$. This yields the 
$\mathbb{C}[X_1,\dots,X_4,X'_1,X'_4]$-module
\begin{equation}
\mathcal{T}_\infty = \frac{\mathbb{C}[X_1,\dots,X_4,X'_1,X'_4,\alpha_1,\alpha_1^{-1},\alpha_2,\alpha_2^{-1}]}{\langle (\alpha_1^{-Q_{1\,i}}\alpha_2^{-Q_{2\,i}}X_i-X_i')_{i=1,4}, (X_2-\alpha_1^{-2}\alpha_2), (X_3-\alpha_1\alpha_2^{-2}),(\alpha_i\alpha^{-1}_i-1) \rangle}. 
\label{T_orb}
\end{equation}
The gauge symmetry of the theory on the right of the defect is broken from $U(1)^R_1\times U(1)^R_2$ to 
\begin{equation}
\left\{\left(e^{\frac{2\pi i n}{3}},e^{-\frac{2\pi i n}{3}}\right)\vert n\in\mathbb{Z}\right\}\cong \mathbb{Z}_3.
\end{equation}
As expected,  $\mathcal{T}_\infty$ is not a finitely generated $\mathbb{C}[X_1,\dots,X_4,X'_1,X'_4]$-module. 
To construct the lift defects, we now proceed by introducing cutoffs to charges of the $U(1)$s unbroken at all the phase boundaries traversed by the chosen path. 
For the transition $(11)\leftrightarrow (01)\leftrightarrow (00)$ we obtain the cutoff
\begin{equation}
Q^L_1+2Q^L_2\leq N
\end{equation}
from the phase boundary $(11)\leftrightarrow (01)$ and the cutoff
\begin{equation}
Q^L_1\leq M
\end{equation}
from $(01)\leftrightarrow (00)$. Instead of the module $\mathcal{T}_\infty$ we consider the submodule
$\mathcal{T}_{N,M}$ generated by only those generators whose charges satisfy these inequalities. 
The choice of cutoff parameters $N$ and $M$ corresponds to the choice of homotopy class of path between the respective phases. 

Note that in $\mathcal{T}_\infty$ there are relations
\begin{equation}
X_3\alpha_1^{-1}\alpha_2^{2}=1\qquad
X_2\alpha_1^{2}\alpha_2^{-1}=1,
\end{equation}
which can be used to write any generator $e$ as 
\begin{equation}\label{eq:relinwindow}
e=\left(X_3\alpha_1^{-1}\alpha_2^{2}\right)^r\left(X_2\alpha_1^{2}\alpha_2^{-1}\right)^s\,e
=X_3^rX_2^s\,\left(\alpha_1^{-1}\alpha_2^{2}\right)^r\left(\alpha_1^{2}\alpha_2^{-1}\right)^se
=:X_3^rX_2^s\,e^\prime.
\end{equation}
Now, the charges under the unbroken $U(1)$s of $\alpha_1^{-1}\alpha_2^{2}$ and $\alpha_1^{2}\alpha_2^{-1}$ are given by
\begin{equation}
\begin{array}{c|cc}
&Q^L_1+2Q^L_2&Q_1^L\\
\hline
\alpha_1^{-1}\alpha_2^{2}&3&-1\\
\alpha_1^{2}\alpha_2^{-1}&0&2
\end{array}
\end{equation}
Thus, any generator of $\mathcal{T}_{N,M}$ can via \eqref{eq:relinwindow} be written as an element of the algebra applied to a generator whose charges lie in the band
\begin{align}
\begin{split}
& N-3< Q^L_1+2Q^L_2\leq N\\
& M-2< Q^L_1\leq M.
\end{split}
\label{eq:bound(11)-(01)}
\end{align}
Thus, $\mathcal{T}_{N,M}$ is generated by generators with charges in this band. In particular it is finitely generated.

Indeed, this matches precisely with the analysis of D-brane transport in \cite{herbst2008}. The charge bands above are the range of charges of D-branes lifted from the orbifold phase to the GLSM by fusion with the respective lift defects. They precisely match with all band restriction rules for this model from \cite{herbst2008} (example (D)):  
\begin{align}
\begin{split}
& -\frac{3}{2}< \frac{\theta_1+2\theta_2}{2\pi}+Q^L_1+2Q^L_2< \frac{3}{2}\\
& -1<\frac{\theta_1}{2\pi}+Q^L_1<1,
\end{split}
\end{align}
where the choice of cutoff parameters $M$ and $N$ corresponds to the choice of homotopy class of paths determined by $\theta_1$ and $\theta_2$.

It is very easy to determine the module $\mathcal{T}_{N,M}$ from the charge bands. 
For $N-M$ even the independent generators satisfying (\ref{eq:bound(11)-(01)}) are 
\begin{equation}
e_1:=\alpha_1^{M}\alpha_2^{\frac{N-M}{2}}\,,\quad
e_2:=\alpha_1^{M}\alpha_2^{\frac{N-M}{2}}\alpha_1^{-1}\,,\quad
e_3:=\alpha_1^{M}\alpha_2^{\frac{N-M}{2}}\alpha_2^{-1}\,.
\end{equation}
They are subject to the relations
\begin{equation}
\begin{aligned}
	& X'_1 e_1= X_1e_2\\
	& X'_1 e_2= X_1X_2e_3\\
	& X'_1 e_3= X_1X_2X_3e_1\\
	& X'_4e_1= X_4e_3\\
	& X'_4e_3= X_3X_4e_2\\
	& X'_4e_2= X_2X_3X_4e_1
\end{aligned}
\end{equation}
The $U(1)_1^L\times U(1)_2^L\times\mathbb{Z}_3^R$-charges of the generators are given by
\begin{equation}
\begin{array}{ll}
e_1:&\left(M,\frac{N-M}{2},\left[\frac{1}{2}(N-3M)\right]_3\right)\\
e_2:&\left(M-1,\frac{N-M}{2},\left[\frac{1}{2}(N-3M)+1\right]_3\right)\\
e_3:&\left(M,\frac{N-M}{2}-1,\left[\frac{1}{2}(N-3M)+2\right]_3\right)
\end{array}
\end{equation}
For $N-M$ odd the independent generators satisfying (\ref{eq:bound(11)-(01)}) are
\begin{equation}
e_1^\prime:=\alpha_1^{M-1}\alpha_2^{\frac{N-M-1}{2}}\,,\quad
e_2^\prime:=\alpha_1^{M-1}\alpha_2^{\frac{N-M-1}{2}}\alpha_1\,,\quad
e_3^\prime:=\alpha_1^{M-1}\alpha_2^{\frac{N-M-1}{2}}\alpha_2\,.
\end{equation}
They satisfy the relations
\begin{equation}
\begin{aligned}
	& X'_1 e_1^\prime= X_1X_2X_3 e_3^\prime\\
	& X'_1 e_3^\prime= X_1X_2e_2^\prime\\
	& X'_1e_2^\prime= X_1e_1^\prime\\
	& X'_4e_1^\prime= X_2X_3X_4e_2^\prime\\
	& X'_4e_2^\prime= X_3X_4e_3^\prime\\
	& X'_4e_3^\prime= X_4e_1^\prime
\end{aligned}
\label{A_2T(0,0)relations}
\end{equation}
and their $U(1)_1^L\times U(1)_2^L\times\mathbb{Z}_3^R$-charges are given by
\begin{equation}\label{eq:chargeseprime}
\begin{array}{ll}
e_1^\prime:&\left(M-1,\frac{N-M-1}{2},\left[\frac{1}{2}(N+1-3M)\right]_3\right)\\
e_2^\prime:&\left(M,\frac{N-M-1}{2},\left[\frac{1}{2}(N+1-3M)+1\right]_3\right)\\
e_3^\prime:&\left(M-1,\frac{N-M+1}{2},\left[\frac{1}{2}(N+1-3M)+2\right]_3\right)
\end{array}
\end{equation}

\subsubsection{Brane Lift to the GLSM}

Lifting D-branes from the orbifold phase to the GLSM is described by fusing the D-branes with the respective lift defects.

In order to illustrate this, we will fuse the lift defects with the fractional D0 branes in the orbifold phase. The latter correspond to the $R'\coloneqq\mathbb{C}[X'_1,X'_4]$-modules
\begin{equation}
\mathbb{C}[X'_1,X'_4]/\langle X'_1,X'_4\rangle\{[n]_3\}. \label{fractional_brane}
\end{equation}
The lift of (\ref{fractional_brane}) is given by the fusion with $\mathcal{T}_{N,M}$ 
\begin{equation}
\mathcal{T}_{N,M}*R'/\langle X'_1,X'_4\rangle\{[n]_3\} =(\mathcal{T}_{N,M}\otimes_{R'} R'/\langle X'_1,X'_4\rangle\{[n]_3\})^{\mathbb{Z}_3},
\end{equation} 
For concreteness let us consider the case  $N-M$ odd and $\frac{1}{2}(N+1-3M)+n=0 \mod 3$.
Then the $\mathbb{Z}_3$-invariant generator in ${\mathcal{T}_{N,M}\otimes_{R'} R'/\langle X'_1,X'_4\rangle\{[n]_3\}}$ is $e_1^\prime\otimes 1$.
Therefore, replacing the variables $X'_1$ and $X'_4$ according to the relations (\ref{A_2T(0,0)relations}) we find the lift
\begin{equation}
\mathcal{T}(N,M)*R'/\langle X'_1,X'_4\rangle\{[n]_3\}={\mathbb C}[X_1, X_2, X_3, X_4]/\langle X_1X_2X_3,X_2X_3X_4 \rangle \left\lbrace\left(M-1,\frac{N-M-1}{2}\right)\right\rbrace.
\end{equation}
This module can be expressed equivalently by its Koszul resolution
\begin{equation}
\resizebox{\textwidth}{!} 
{
$
R  \otimes \bigwedge\nolimits^2 V^\vee\left\lbrace\left(M-1,\frac{N-M-1}{2}\right)\right\rbrace\overset{d_1}{\rightarrow} R\otimes(V_4^\vee\left\lbrace\left(M-1,\frac{N-M+1}{2}\right)\right\rbrace\oplus V_1^\vee\left\lbrace\left(M,\frac{N-M-1}{2}\right)\right\rbrace\overset{d_2}{\rightarrow}  R\left\lbrace\left(M-1,\frac{N-M-1}{2}\right)\right\rbrace,
$
}
\end{equation}
where $V\coloneqq \text{span}\{\pi_1,\pi_4\}\eqqcolon V_1\oplus V_4$ is a two-dimensional vector space, $R\coloneqq\bC [X_1,X_2,X_3,X_4]$ and
 \begin{equation}
d_1 = 
	\begin{pmatrix}
	X_1\\
	-X_4
	\end{pmatrix},\ 
d_2 =
	\begin{pmatrix}
	X_2X_3X_4 & X_1X_2X_3
	\end{pmatrix}.
\end{equation} 

Indeed, we may arrive at the same result by taking a slightly different route, first determining the resolution of the fractional brane (\ref{fractional_brane}) and then performing the fusion with $\mathcal{T}_{N,M}$.
The fractional brane (\ref{fractional_brane}) can be represented by the Koszul resolution 
\begin{equation}
A^\bullet\coloneqq R'\otimes \bigwedge\nolimits^2 V^\vee\{[n]_3\} \rightarrow R'\otimes(V_4^\vee\{[n+1]_3\}\oplus V_1^\vee\{[n-1]_3\})\rightarrow R'\{[n]_3\},
\end{equation}
with differential $d\coloneqq\sum_i X'_i\pi_i\lrcorner$.
To compute the lift
\begin{equation}
\mathcal{T}_{N,M}*A^\bullet =(\mathcal{T}_{N,M}\otimes_{R'} A^\bullet)^{\mathbb{Z}_3} 
\end{equation} 
of (\ref{fractional_brane}) we have to determine the $\bZ_3$-invariant part of $\mathcal{T}_{N,M}\otimes_{R'} R'\{[n]_3\}$. This can be accomplished by using the charges \eqref{eq:chargeseprime} of the generators $e_i^\prime$ of $\mathcal{T}_{N,M}$. One obtains
\begin{equation}
\resizebox{\textwidth}{!} 
{
$
R  \otimes \bigwedge\nolimits^2 V^\vee\left\lbrace\left(M-1,\frac{N-M-1}{2}\right)\right\rbrace\overset{d_1}{\rightarrow} R\otimes(V_4^\vee\left\lbrace\left(M-1,\frac{N-M+1}{2}\right)\right\rbrace\oplus V_1^\vee\left\lbrace\left(M,\frac{N-M-1}{2}\right)\right\rbrace\overset{d_2}{\rightarrow}  R\left\lbrace\left(M-1,\frac{N-M-1}{2}\right)\right\rbrace.
$
}
\end{equation}
Here 
\begin{equation}
d_1 = 
	\begin{pmatrix}
	A\\
	-B
	\end{pmatrix},\ 
d_2 =
	\begin{pmatrix}
	C & D
	\end{pmatrix}
\end{equation}
with
\begin{equation}
A=X_1\,,\quad B=X_4\,,\quad C=X_2X_3X_4\,,\quad D=X_1X_2X_3\,,
\end{equation}
which has been obtained by replacing $X'_i$ variables with the $X_i$ variables according to the relations (\ref{A_2T(0,0)relations}).

Of course, one can perform the calculations also for the other fractional branes. The lifts for these cases are given by
\begin{equation}
\resizebox{\textwidth}{!} 
{
$
R  \otimes \bigwedge\nolimits^2 V^\vee\left\lbrace\left(M-1,\frac{N-M+1}{2}\right)\right\rbrace\overset{d_1}{\rightarrow} R\otimes(V_4^\vee\left\lbrace\left(M,\frac{N-M-1}{2}\right)\right\rbrace\oplus V_1^\vee\left\lbrace\left(M-1,\frac{N-M-1}{2}\right)\right\rbrace\overset{d_2}{\rightarrow}  R\left\lbrace\left(M-1,\frac{N-M+1}{2}\right)\right\rbrace,
$
}
\end{equation}
for $\frac{1}{2}(N+1-3M)+n=-1 \ \text{mod}  \ 3$ where now
\begin{equation}
A=X_1X_2X_3\,,\quad B=X_3X_4\,,\quad C=X_4\,,\quad D=X_1X_2\,,
\end{equation}
and
\begin{equation}
\resizebox{\textwidth}{!} 
{
$
R  \otimes \bigwedge\nolimits^2 V^\vee\left\lbrace\left(M,\frac{N-M-1}{2}\right)\right\rbrace\overset{d_1}{\rightarrow} R\otimes(V_4^\vee\left\lbrace\left(M-1,\frac{N-M-1}{2}\right)\right\rbrace\oplus V_1^\vee\left\lbrace\left(M-1,\frac{N-M+1}{2}\right)\right\rbrace\overset{d_2}{\rightarrow}  R\left\lbrace\left(M,\frac{N-M-1}{2}\right)\right\rbrace,
$
}
\end{equation}
for $\frac{1}{2}(N+1-3M)+n=-2 \ \text{mod}  \ 3$ with
\begin{equation}
A=X_1X_2\,,\quad B=X_2X_3X_4\,,\quad C=X_3X_4\,,\quad D=X_1\,.
\end{equation}
This agrees with the results in 
8.4.2 (D) of \cite{herbst2008}, where a lift of the fractional branes was determined for the homotopy class of paths given by 
$\theta_1=-\pi,\ \theta_2=-\frac{\pi}{2}$. This corresponds to the choice of our cutoff parameters $(N,M)=(2,1)$.

\subsection{A Two Parameter Model with $\mathbb{C}^5/\mathbb{Z}_8$-orbifold phase}

Next we will apply our construction to another two parameter model, which is defined by the following data
\begin{equation}
\text{GLSM}_{\mathbb{Z}_8-\textup{orb}}=(U(1)_1\times U(1)_2, V,(r,\theta),W=0).
\end{equation}
Here the representation $V$, can be read off from the charge assignment of the chiral matter content given by table~\ref{U(1)xU(1)} below.
(This is example (C) in \cite{herbst2008}.)

\begin{table}[H]
\centering
	\begin{tabular}{c|ccccccccc}
	& $X_0$ & $X_1$ & $X_2$ & $X_3$ & $X_4$ & $X_5$ & $X_6$ \\
	\hline
	$Q_1$ & -4 & 0 & 0 & 1 & 1 & 1 & 1\\
	$Q_2$ & 0 & 1 & 1 & 0 & 0 & 0 & -2\\
	\end{tabular}
\caption{Matter content of the two parameter model. $Q_j$ denotes the $U(1)_j$-charge of the field $X_i$.}
\label{U(1)xU(1)}
\end{table}
\noindent
The phase structure of this model is similar to the one of the $A_2$-model. It exhibits an orbifold phase for $r_1,\ r_2\rightarrow -\infty$.
In this phase the fields $X_0$ and $X_6$ aquire a vev and the low energy theory is the orbifold theory $\mathbb{C}_5/\mathbb{Z}_8$. We call it phase III. It also features a geometric, or large volume phase (I) in the regime $r_1,\ r_2\rightarrow \infty$. This phase has an effective description  by a non-linear sigma model on the total space of the line bundle ${\mathcal O}(-8)$ over the weighted projective space $\mathbb{P}_{(11222)}$. Beyond these, there are two mixed phases (II, IV). 

The phase diagram of $\text{GLSM}_{\mathbb{Z}_8-\textup{orb}}$ is depicted in Figure~\ref{A_U1xU1Phases} below.

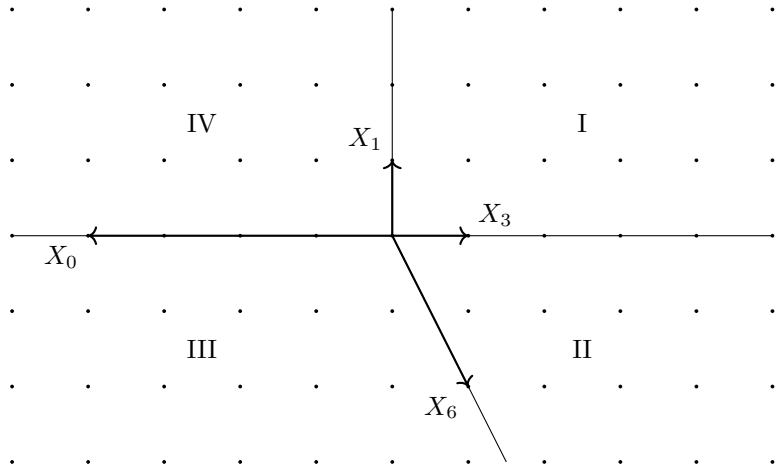
\begin{figure}[h]
\center
\begin{tikzpicture}
\foreach \x in {-5,-4,...,5} {
\foreach \y in {-3,-2,...,3} {
\fill[color=black] (\x,\y) circle (0.25mm);
}
}
\draw[thick,->] (0,0) -- (1,0) node[anchor=south west] {$X_3$};
\draw[thick,->] (0,0) -- (-4,0) node[anchor=north east] {$X_0$};
\draw[thick,->] (0,0) -- (1,-2) node[anchor=north east] {$X_6$};
\draw[thick,->] (0,0) -- (0,1) node[anchor=south east] {$X_1$};
\draw[thick] (2.5,1.5) node[] {I};
\draw[thick] (-2.5,-1.5) node[] {III};
\draw[thick] (2.5,-1.5) node[] {II};
\draw[thick] (-2.5,1.5) node[] {IV};
\draw[thin,-] (0,0) -- (0,3);
\draw[thin,-] (0,0) -- (5,0) ;
\draw[thin,-] (0,0) -- (-5,0);
\draw[thin,-] (0,0) -- (1.5,-3);
\end{tikzpicture}
\caption{Phase diagram of $\text{GLSM}_{\mathbb{Z}_8-\textup{orb}}$. Here, I is the large volume phase, III is the orbifold phases and IV and II are mixed phases.}
\label{A_U1xU1Phases}
\end{figure}
\noindent
The locations of the phase boundaries and the respective unbroken gauge groups are given by
\begin{center}
\begin{tabular}{ccc}
phase boundary & location & unbroken gauge groups\\
\hline
$(\text{III})\leftrightarrow (\text{IV})$&$\text{Cone}_{\{0\}}$&$\mathbb{Z}_4\times U(1)_2$\\
$(\text{IV})\leftrightarrow (\text{I})$&$\text{Cone}_{\{1\}}$&$U(1)_1$\\
$(\text{III})\leftrightarrow (\text{II})$&$\text{Cone}_{\{6\}}$&$\{(g^2,g)\,|\,g\in U(1)\}$\\
$(\text{II})\leftrightarrow (\text{I})$&$\text{Cone}_{\{3\}}$&$U(1)_2$
\end{tabular}
\end{center}

\subsubsection{GLSM$_{\mathbb{Z}_8-\textup{orb}}$ Identity Defect}

The identity defect of GLSM$_{\mathbb{Z}_8-\textup{orb}}$ is associated to the $\mathbb{C}[X_0,\dots,X_6,X'_0,\dots,X'_6]$-module
\begin{equation}
\mathcal{I}_{\text{GLSM}_{\mathbb{Z}_8-\textup{orb}}}\coloneqq\mathbb{C}[X_0,\dots,X_6,X'_0,\dots,X'_6,\alpha_1,\alpha_1^{-1},\alpha_2,\alpha_2^{-1}]/\langle (\alpha_1^{-Q_{1\,i}}\alpha_2^{-Q_{2\,i}}X_i-X'_i),(\alpha_i\alpha_i^{-1}-1) \rangle,
\end{equation}
where an additional copy of the fields of GLSM$_{\mathbb{Z}_8-\textup{orb}}$ as well as auxiliary fields $\alpha_i$ and $\alpha_i^{-1}$ were introduced. 
The gauge-charges of the various fields are given in Table~\ref{U1xU1GLSMId} below.

\begin{table}[H]
\centering
	\begin{tabular}{c|*{12}{>{\centering\arraybackslash}p{.05\linewidth}}}
	& $X_0$ & $X_{1,2}$ & $X_{3,4,5}$ & $X_6$ & $X'_0$ & $X'_{1,2}$ & $X'_{3,4,5}$ & $X'_6$ & $\alpha_1$ & $\alpha^{-1}_1$ & $\alpha_2$ & $\alpha^{-1}_2$\\
	\hline
	$Q_1^L$ & -4 & 0 & 1 & 1 & 0 & 0 & 0 & 0 & 1 & -1 & 0 & 0\\
	$Q_2^L$ & 0 & 1 & 0 & -2 & 0 & 0 & 0 & 0 & 0  & 0 & 1 & -1\\
	$Q_1^R$ & 0 & 0 & 0 & 0 & -4 & 0 & 1 & 1 & -1  & 1 & 0 & 0\\		
	$Q_2^R$ & 0 & 0 & 0 & 0 & 0 & 1 & 0 & -2 & 0  & 0 & -1 & 1\\
	\end{tabular}
\caption{Fields of the identity defect of GLSM$_{\mathbb{Z}_8-\textup{orb}}$. $Q_j^L$ and $Q_j^R$ denote the charge under the gauge groups $U(1)^L_j$, respectively $U(1)^R_j$ on the left and right of the defect.}
\label{U1xU1GLSMId}
\end{table}
\noindent

\subsubsection{Orbifold Lift}

Starting from $\mathcal{I}_{\text{GLSM}_{\mathbb{Z}_8-\textup{orb}}}$ we construct the lift from the orbifold phase to $\text{GLSM}_{\mathbb{Z}_8-\textup{orb}}$ by setting $X'_0=X'_6=1$.
This yields
\begin{equation}
\mathcal{T}_\infty\coloneqq\mathbb{C}[X_0,\dots,X_6,X'_1,\dots,X'_5,\alpha_i^{\pm 1}]/\langle (X_0-\alpha_1^{-4}),(\alpha_1^{-Q_{1\,i}}\alpha_2^{-Q_{2\,i}}X_i-X'_i),(X_6-\alpha_1\alpha_2^{-2}),(\alpha_i\alpha_i^{-1}-1) \rangle.
\label{T_orb2}
\end{equation}
The gauge symmetry on the right of the defect is broken to 
$U(1)_{1}^R\times U(1)_{2}^R\rightarrow \{(g^2,g)\vert g^8=1\}\cong\mathbb{Z}_8$.

As before, we introduce charge cutoffs in this module to obtain the modules associated to the lift defects.  Let us first consider paths $(\text{III})\leftrightarrow (\text{IV})\leftrightarrow (\text{I})$ from the orbifold to the large volume phase transversing phase IV. The unbroken $U(1)$s on the phase boundaries traversed are $U(1)_2$ and $U(1)_1$ respectively. Thus, we have to impose charge cutoffs
\begin{equation}
Q^L_2\leq N
\end{equation}
and
\begin{equation}
Q^L_1\leq M
\end{equation}
for a choice of integers $M$ and $N$. The submodule $\mathcal{T}_{N,M}^{\text{III-IV-I}}\subset\mathcal{T}_\infty$ generated by generators whose charges satisfy these inequalities corresponds to the desired lift defect. Here the choice of cutoff parameters corresponds to a choice of homotopy class of paths from phase III to phase I traversing phase IV.

As in the $A_2$-example discussed above, the relations in $\mathcal{T}_\infty$ render  $\mathcal{T}^{\text{III-IV-I}}_{N,M}$ finitely generated. More precisely, one has relations
\begin{equation}\label{eq:z8rel}
X_6\alpha_1^{-1}\alpha_2^{2}=1\;\text{and}\;
X_0\alpha_1^{4}=1.
\end{equation}
The charges under the unbroken $U(1)$s of $\alpha_1^{-1}\alpha_2^{2}$ and $\alpha_1^{4}$ are given by
\[
\begin{array}{c|cc}
&Q^L_2&Q_1^L\\
\hline
\alpha_1^{-1}\alpha_2^{2}&2&-1\\
\alpha_1^{4}&0&4
\end{array}
\]
Hence the charges of independent generators of the truncated module $\mathcal{T}_{N,M}^{\text{III-IV-I}}$ lie in the bands
\begin{align}
\begin{split}
& N-2< Q^L_2\leq N\\
& M-4< Q^L_1\leq M.
\end{split}
\label{eq:bound(III)-(IV)-(I)}
\end{align}
It is not difficult to determine all the generators of $\mathcal{T}_{N,M}^{\text{III-IV-I}}$. 
The independent generators satisfying (\ref{eq:bound(III)-(IV)-(I)}) are given by 
\begin{equation}
\alpha_1^{M-3}\alpha_2^{N-1} \{e_1\coloneqq 1,\ e_2\coloneqq\alpha_1^3\alpha_2,\ e_3\coloneqq\alpha_1^3,\ e_4\coloneqq\alpha_1^2\alpha_2\,\ e_5\coloneqq\alpha_1^2,\ e_6\coloneqq\alpha_1\alpha_2,\ e_7\coloneqq\alpha_1,\ e_8\coloneqq\alpha_2 \}.
\end{equation}

As for paths $(\text{III})\leftrightarrow (\text{II})\leftrightarrow (\text{I})$ from phase III to phase I traversing the other intermediate phase II, the respective unbroken $U(1)$s are given by $\{(g^2,g)\,|\,g\in U(1)\}$ and $U(1)_2$. Therefore, the corresponding cutoffs read
\begin{equation}
2Q^L_1+Q^L_2\leq N\;\text{and}\; Q^L_2\leq M.
\end{equation}
By virtue of the relations \eqref{eq:z8rel}, one again finds that the truncated modules $\mathcal{T}_{N,M}^{\text{III-II-I}}$ are finitely generated. 
The charges of $\alpha_1^{-1}\alpha_2^{2}$ and $\alpha_1^{4}$ under the unbroken $U(1)$s  are given by
\[
\begin{array}{c|cc}
&2Q^L_1+Q^L_2&Q_2^L\\
\hline
\alpha_1^{-1}\alpha_2^{2}&0&2\\
\alpha_1^{4}&8&0
\end{array}
\]
Therefore, the independent generators of $\mathcal{T}_{N,M}^{\text{III-II-I}}$ are the ones whose charges lie in the band
\begin{align}
\begin{split}
& N-8< 2Q^L_1+Q^L_2\leq N\\
& M-2< Q^L_2\leq M.
\end{split}
\label{eq:bound(III)-(II)-(I)}
\end{align}
The independent generators of $\mathcal{T}_{N,M}^{\text{III-II-I}}$
are given by
\begin{equation}
\alpha_1^{\frac{N-M-7}{2}}\alpha_2^{M-1}\{e'_1\coloneqq \alpha_1^4,\ e'_2\coloneqq\alpha_1^3\alpha_2,\ e'_3\coloneqq\alpha_1^3,\ e'_4\coloneqq\alpha_1^2\alpha_2\,\ e'_5\coloneqq\alpha_1^2,\ e'_6\coloneqq\alpha_1\alpha_2,\ e'_7\coloneqq\alpha_1,\ e'_8\coloneqq\alpha_2 \}.
\end{equation}
for $N-M$ odd and 
\begin{equation}
\alpha_1^{\frac{N-M-6}{2}}\alpha_2^{M-1} \{e_1,\ e_2,\ e_3,\ e_4\,\ e_5,\ e_6,\ e_7,\ e_8 \}.
\end{equation}
for $N-M$ even.

As for the $A_2$-model we find complete agreement with \cite{herbst2008}. The charge bands 
\eqref{eq:bound(III)-(IV)-(I)} and \eqref{eq:bound(III)-(II)-(I)} of the lift defects associated to paths $(\text{III})\leftrightarrow(\text{IV})\leftrightarrow(\text{I})$, respectively $(\text{III})\leftrightarrow(\text{II})\leftrightarrow(\text{I})$ match with the respective band restriction rules, c.f.~Section 4.4 (C) in \cite{herbst2008}:
\begin{align}
\begin{array}{cc}
(\text{III})\leftrightarrow(\text{IV})\leftrightarrow(\text{I})&(\text{III})\leftrightarrow(\text{II})\leftrightarrow(\text{I})\\\hline
 -1< \frac{\theta_2}{2\pi}+2Q^L_2< 1&-4<\frac{2\theta_1+\theta_2}{2\pi}+2Q^L_1+Q^L_2<4\\
 -2<\frac{\theta_1}{2\pi}+Q^L_1<2 &-1< \frac{\theta_2}{2\pi}+Q^L_2< 1.
\end{array}
\end{align}
To be more concrete,  let us briefly list the charges of the modules for specific choices of cutoff parameters. The $U(1)_1^L\times U(1)_2^R\times\mathbb{Z}_8^R$ charges of the generators of $\mathcal{T}_{\text{III-IV-I}}(1,3)$ as well as $\mathcal{T}_{\text{III-II-I}}(7,1)$ are given by
\[\begin{array}{cccccccc}
e_1&e_2&e_3&e_4&e_5&e_6&e_7&e_8\\
\hline
(0,0,[0]_8)&(3,1,[1]_8)&(3,0,[2]_8)&(2,1,[3]_8)&(2,0,[4]_8)&(1,1,[5]_8)&(1,0,[6]_8)&(0,1,[7]_8)
\end{array}
\]
whereas the generators of $\mathcal{T}_{\text{III-II-I}}(8,1)$ have charges
\[
\begin{array}{cccccccc}
e_1&e_2&e_3&e_4&e_5&e_6&e_7&e_8\\
\hline
(4,0,[0]_8)&(3,1,[1]_8)&(3,0,[2]_8)&(2,1,[3]_8)&(2,0,[4]_8)&(1,1,[5]_8)&(1,0,[6]_8)&(0,1,[7]_8)
\end{array}
\]
This agrees with the band restrictions 
\begin{align}
\begin{split}
\mathfrak{C}^{w_0}_{\text{III},\text{II}}\cap\mathfrak{C}^{w'}_{\text{II},\text{I}}=\{(0,1),(1,0),(1,1),(2,0),(2,1),(3,0),(3,1),(4,0)\},\\
\mathfrak{C}^{w_1}_{\text{III},\text{II}}\cap\mathfrak{C}^{w'}_{\text{II},\text{I}}=\mathfrak{C}^{w'}_{\text{III},\text{IV}}\cap\mathfrak{C}^{w''}_{\text{IV},\text{I}}=\{(0,0),(0,1),(1,0),(1,1),(2,0),(2,1),(3,0),(3,1)\}
\end{split}
\end{align}
determined in \cite{herbst2008} Section 8.4.2.~(C), where the roman numerals label the phases and 
\begin{align}
\begin{split}
& w_0:\ -10\pi<2\theta_1+\theta_2<-8\pi \\
& w_1:\ -8\pi<2\theta_1+\theta_2<-6\pi\\
& w':\ -2\pi<\theta_2<0.
\end{split}
\end{align}

\subsubsection{Brane Lift to the GLSM}

In \cite{herbst2008} Section 8.4.2. (C) a lift of the fractional brane 
\begin{equation}
\mathbb{C}[X_1^\prime,\dots,X_5^\prime]/\langle X_1^\prime,\dots,X_5^\prime \rangle\{[0]_8\}
\end{equation}
\noindent
from the $\mathbb{C}^5/\mathbb{Z}_8$-phase to the GLSM is performed.
For comparisons sake we briefly outline the lift of this brane via fusion with the transition defects constructed above.
The fractional brane can be represented by the $R^\prime\coloneqq\mathbb{C}[X_1^\prime,\dots,X_5^\prime]$-free resolution

\adjustbox{scale=.7,center}
{
\begin{tikzcd}
                  &                                                               &                                                               & {\Lambda^3V^\vee\otimes R^\prime\{[2]_8\}} \arrow[rd, "Y"]                 &                                                          &                                      &                             &   \\
                  &                                                               & {\Lambda^4V^\vee\otimes R^\prime\{[1]_8\}} \arrow[ru, "X"] \arrow[rd, "Y"] &                                                         & {\Lambda^2V^\vee\otimes R^\prime\{[4]_8\}} \arrow[rd, "Y"]            &                                      &                             &   \\
0 \arrow[r] & {\Lambda^5V^\vee\otimes R^\prime\{[0]_8\}} \arrow[rd, "Y"] \arrow[ru, "X"] &                                                         & {\Lambda^3V^\vee\otimes R^\prime\{[3]_8\}} \arrow[ru, "X"] \arrow[rd, "Y"] &                                                    & {V^\vee\otimes R^\prime\{[6]_8\}} \arrow[rd, "Y"] &                             &   \\
                  &                                                               & {\Lambda^4V^\vee\otimes R^\prime\{[2]_8\}} \arrow[ru, "X"] \arrow[rd, "Y"] &                                                         & {\Lambda^2V^\vee\otimes R^\prime\{[5]_8\}} \arrow[rd] \arrow[ru, "X"] &                                & {R^\prime\{[0]_8\}} \arrow[r] & 0, \\
                  &                                                               &                                                               & {\Lambda^3V^\vee\otimes R^\prime\{[4]_8\}} \arrow[ru, "X"] \arrow[rd, "Y"] &                                                    & {V^\vee\otimes R^\prime\{[7]_8\}} \arrow[ru, "X"] &                             &   \\
                  &                                                               &                                                               &                                                               & {\Lambda^2V^\vee\otimes R^\prime\{[6]_8\}} \arrow[ru, "X"]            &                                      &                             &  
\end{tikzcd}
}

\noindent
where $V\coloneqq\text{span}\{\pi_1,\dots,\pi_5\},\ X=\sum\limits_{i=1}^2X_i^\prime\pi_i\iprod$ is a five-dimensional vector space and  $Y=\sum\limits_{i=3}^5X_i^\prime\pi_i\iprod$.
We lift this brane via fusion with $\mathcal{T}_{\text{III-IV-I}}(1,3)$, this yields the band restricted brane

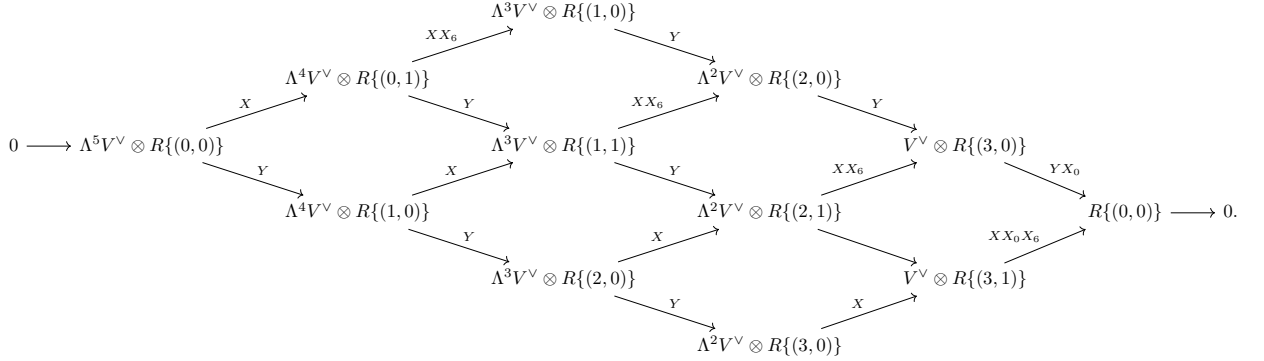
\adjustbox{scale=.7,center}
{
\begin{tikzcd}
                  &                                                                 &                                                                    & {\Lambda^3V^\vee\otimes R\{(1,0)\}} \arrow[rd, "Y"]                    &                                                               &                                              &                               &    \\
                  &                                                                 & {\Lambda^4V^\vee\otimes R\{(0,1)\}} \arrow[ru, "XX_6"] \arrow[rd, "Y"] &                                                              & {\Lambda^2V^\vee\otimes R\{(2,0)\}} \arrow[rd, "Y"]               &                                              &                               &    \\
0 \arrow[r] & {\Lambda^5V^\vee\otimes R\{(0,0)\}} \arrow[rd, "Y"] \arrow[ru, "X"] &                                                              & {\Lambda^3V^\vee\otimes R\{(1,1)\}} \arrow[ru, "XX_6"] \arrow[rd, "Y"] &                                                         & {V^\vee\otimes R\{(3,0)\}} \arrow[rd, "YX_0"]    &                               &    \\
                  &                                                                 & {\Lambda^4V^\vee\otimes R\{(1,0)\}} \arrow[ru, "X"] \arrow[rd, "Y"]    &                                                              & {\Lambda^2V^\vee\otimes R\{(2,1)\}} \arrow[rd] \arrow[ru, "XX_6"] &                                        & {R\{(0,0)\}} \arrow[r] & 0. \\
                  &                                                                 &                                                                    & {\Lambda^3V^\vee\otimes R\{(2,0)\}} \arrow[ru, "X"] \arrow[rd, "Y"]    &                                                         & {V^\vee\otimes R\{(3,1)\}} \arrow[ru, "XX_0X_6"] &                               &    \\
                  &                                                                 &                                                                    &                                                                    & {\Lambda^2V^\vee\otimes R\{(3,0)\}} \arrow[ru, "X"]               &                                              &                               &   
\end{tikzcd}
}
\noindent
Just as for the $A_2$-model considered in the previous section this agrees with the respective lift in \cite{herbst2008}.
\subsection{The $A_{N-1}$-Model}\label{sec:anmodel}

Next, we will generalize the treatment of the $A_2$-model in Section~\ref{sec:a2} to the GLSMs associated to $A_{N-1}$-singularities for general $N$. The GLSM is given by the data
\begin{equation}
	\text{GLSM}_{A_{N-1}}\coloneqq(U(1)^{N-1},\ V,\ (r,\theta),\ W=0),
\end{equation}
where the representation $V$, can be read off from the charge assignment of the chiral matter content given by table~\ref{A_N-1Matter} below.

\begin{table}[H]
\centering
	\begin{tabular}{c|ccccccccc}
	& $X_1$ & $X_2$ & $X_3$ & $X_4$ & $\dots$ & $X_{N-1}$ & $X_N$ & $´X_{N+1}$\\
	\hline
	$Q_1$ & 1 & -2 & 1 & 0 & $\dots\ $ 0 & 0 & 0 & 0\\
	$Q_2$ & 0 & 1 & -2 & 1 & $\dots\ $ 0 & 0 & 0 & 0\\
	$\vdots$ & $\vdots$ & & & & & & & $\vdots$ \\
	$Q_{N-2}$ & 0 & 0 & 0 & 0 & $\dots\ $ 1 & -2 & 1 & 0\\
	$Q_{N-1}$ & 0 & 0 & 0 & 0 & $\dots\ $ 0 & 1 & -2 & 1\\
	\end{tabular}
\caption{Matter content of the $A_{N-1}$-model. $Q_j$ denotes the charge under $U(1)_j$.}
\label{A_N-1Matter}
\end{table}
\noindent
Just as the $A_2$-model the $A_{N-1}$-model for general $N$ also exhibits an orbifold phase (${r_1,\dots,r_{N-1}\rightarrow -\infty}$). In this phase, the fields $X_2,\ldots,X_{N}$ aquire a non-trivial vacuum expecation value, and 
the low energy theory is given by the orbifold model $\mathbb{C}^2/ \mathbb{Z}_N$ with fields $X_1$ and $X_{N+1}$ whose $\mathbb{Z}_N$-charges are $1$ and $-1$ respectively. On the opposite side, in the regime ${r_1,\dots,r_{N-1}\rightarrow \infty}$ the theory is effectively described by a non-linear sigma model on the resolution of the $A_{N-1}$-singularity. This is the geometric or large volume phase. In total, the model has $2^{N-1}$ phases associated to the various partial resolutions of the singularity. The $A_{N-1}$ singularity has an exceptional divisor of $N-1$ 2-spheres intersecting according to the $A_{N-1}$-Dynkin diagram, and a partial resolution is determined by which of these 2-spheres is blown up. We label the different phases by an $N-1$-tuple, whose $i$th entry is either $0$ or $1$ according to whether the $i$th 2-sphere in the exceptional divisor is blown up, or not. In particular $(00\dots0)$ corresponds to the full resolution of the singularity and therefore to the large volume phase, whereas $(11\ldots 1)$ denotes to the unresolved singularity and hence the orbifold phase.
For more details on the $A_{N-1}$-model, see e.g.~example (D) in \cite{herbst2008}.

\subsubsection{GLSM$_{A_{N-1}}$ Identity Defect}

To write down the identity defect of $\text{GLSM}_{A_{N-1}}$ we introduce a pair of auxiliary defect fields for each $U(1)$-factor of the gauge group denoted $\alpha_i,\ \alpha_i^{-1}$, $\ i=1,\dots,N-1$.
The fields $\alpha_i$ carry $U(1)_i^L\times U(1)_i^R$-charge $(1,-1)$ and are uncharged under all other $U(1)$s whereas the fields  $\alpha_i^{-1}$ carry $U(1)_i^L\times U(1)_i^R$-charge $(-1,1)$ and are likewise uncharged with respect to the other $U(1)$s.
The $\text{GLSM}_{A_{N-1}}$ identity defect is associated to the 
$\mathbb{C}[X_1,\dots ,X_{N+1},X'_1,\dots ,X'_{N+1}]$-module
\begin{equation}
\mathcal{I}_{\text{GLSM}_{A_{N-1}}}\coloneqq\frac{\mathbb{C}[X_1,\dots ,X_{N+1},X'_1,\dots ,X'_{N+1},\alpha_1,\dots,\alpha_{N-1},\alpha_1^{-1},\dots,\alpha_{N-1}^{-1}]}{\langle (\alpha_1^{-Q_{1\,i}}\dots\alpha_{N-1}^{-Q_{N-1\,i}}X_i-X'_i)_{i=1,\dots,N+1},(\alpha_i\alpha_i^{-1}-1)_{i=1,\dots,N-1}\rangle},
\end{equation}
where the fields $X_1,\dots,X_{N-1}$ and $X'_1,\dots,X'_{N-1}$ are the bulk fields of the GLSM on the left, respectively right of the defect.

\subsubsection{Orbifold Lift}

To construct the defects lifting the orbifold phase to $\text{GLSM}_{A_{N-1}}$ for arbitrary $N$ we proceed in an analogous fashion to the case $N=3$ discussed in Section~\ref{sec:a2} above.
We start out by descending to the orbifold phase on the right side of the GLSM identity defect. This is accomplished by setting $X'_2=\dots=X'_N=1$ in $\mathcal{I}_{\text{GLSM}_{A_{N-1}}}$.
This yields the $\mathbb{C}[X_1,\dots,X_{N+1},X'_1,X'_{N+1}]$-module
\begin{equation}
\mathcal{T}_\infty\coloneqq\frac{\mathbb{C}[X_1,\dots ,X_{N+1},X'_1 ,X'_{N+1},\alpha_1,\dots,\alpha_{N-1},\alpha_1^{-1},\dots,\alpha_{N-1}^{-1}]}{\langle (\alpha_1^{-1}X_1-X'_1),(\alpha_{N-1}^{-1}X_{N+1}-X'_{N+1}),(\alpha_1^{-Q_{1\,i}}\dots\alpha_{N-1}^{-Q_{N-1\,i}}X_i-1)_{i=2,\dots,N},(\alpha_i\alpha_i^{-1}-1)_{i=1,\dots,N-1}\rangle}.
\label{eq:TinftyAN-1}
\end{equation}
The gauge group on the right side of the defect is broken to
\begin{equation}
U(1)_1^R\times\dots\times U(1)_{N-1}^R\longrightarrow \{(g,g^2,\dots,g^{N-1})\vert g\in\mathbb{Z}_N\}.
\end{equation}

\subsubsection*{The Case $N=4$}

We proceed by first discussing the $A_3$ case and then generalizing to arbitrary $A_{N-1}$. 
The phase structure of the $A_3$-model is depicted in Figure \ref{A_3Phases} below.
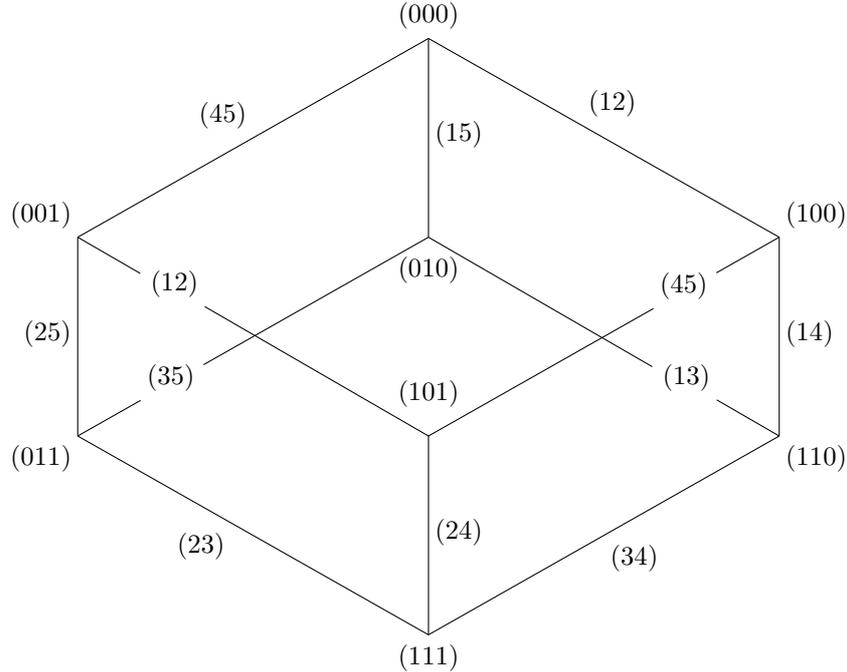
\begin{figure}[h]
\center
\begin{tikzpicture}[x=0.75pt,y=0.75pt,yscale=-1,xscale=1]
 
\draw    (347,151) -- (347,199.26) -- (347,251) ;
\draw    (522,151) -- (347,251) ;
\draw    (347,251) -- (248.62,194.78) -- (172,151) ;
\draw    (172,51) -- (172,99.26) -- (172,151) ;
\draw    (522,51) -- (522,99.26) -- (522,151) ;
\draw    (347,-49) -- (347,-0.74) -- (347,51) ;
\draw    (522,51) -- (423.62,-5.22) -- (347,-49) ;
\draw    (347,-49) -- (172,51) ;
\draw    (347,51) -- (234.78,115.12) ;
\draw    (347,151) -- (235.36,85.88) ;
\draw    (203.34,133.09) -- (172,151) ;
\draw    (203.18,69.19) -- (172,51) ;
\draw    (458.64,116.12) -- (347,51) ;
\draw    (522,151) -- (490.82,132.81) ;
\draw    (459.22,86.88) -- (347,151) ;
\draw    (522,51) -- (490.66,68.91) ;

\draw (347,254.4) node [anchor=north] [inner sep=0.75pt]    {$(111)$};
\draw (170,154.4) node [anchor=north east] [inner sep=0.75pt]    {$(011)$};
\draw (347,137.6) node [anchor=south] [inner sep=0.75pt]    {$(101)$};
\draw (524,154.4) node [anchor=north west][inner sep=0.75pt]    {$(110)$};
\draw (170,47.6) node [anchor=south east] [inner sep=0.75pt]    {$(001)$};
\draw (347,59.4) node [anchor=north] [inner sep=0.75pt]    {$(010)$};
\draw (524,47.6) node [anchor=south west] [inner sep=0.75pt]    {$(100)$};
\draw (347,-52.4) node [anchor=south] [inner sep=0.75pt]    {$(000)$};
\draw (246.62,198.18) node [anchor=north east] [inner sep=0.75pt]    {$(23)$};
\draw (170,99.26) node [anchor=east] [inner sep=0.75pt]    {$(25)$};
\draw (425.62,-8.62) node [anchor=south west] [inner sep=0.75pt]    {$(12)$};
\draw (524,99.26) node [anchor=west] [inner sep=0.75pt]    {$(14)$};
\draw (436.5,204.4) node [anchor=north west][inner sep=0.75pt]    {$(34)$};
\draw (257.5,-2.4) node [anchor=south east] [inner sep=0.75pt]    {$(45)$};
\draw (349,199.26) node [anchor=west] [inner sep=0.75pt]    {$(24)$};
\draw (349,-0.74) node [anchor=west] [inner sep=0.75pt]    {$(15)$};
\draw (233.36,82.48) node [anchor=south east] [inner sep=0.75pt]    {$(12)$};
\draw (205.34,129.69) node [anchor=south west] [inner sep=0.75pt]    {$(35)$};
\draw (488.82,129.41) node [anchor=south east] [inner sep=0.75pt]    {$(13)$};
\draw (461.22,83.48) node [anchor=south west] [inner sep=0.75pt]    {$(45)$};
\end{tikzpicture}
\caption{Phase diagram of $\text{GLSM}_{A_3}$. Here, phases are depicted by vertices, and phase boundaries by edges between them. The orbifold phase is labeled by $(111)$ and the large volume phase where the singularity is fully resolved is labeled by $(000)$. The phase boundaries $(ij)$ are located at $\text{Cone}_{\{i,j\}}$.}
\label{A_3Phases}
\end{figure}
To exemplify our construction, we will 
derive the lift defects associated to paths from the orbifold phase $(111)$ to the geometric phase $(000)$ along two different classes of paths:
 $(111)\leftrightarrow (011)\leftrightarrow (001)\leftrightarrow (000)$ and $(111)\leftrightarrow (101)\leftrightarrow (001)\leftrightarrow (000)$.

We start out with the first one: $(111)\leftrightarrow (011)\leftrightarrow (001)\leftrightarrow (000)$.
The relevant unbroken $U(1)$s at the traversed phase boundaries are given by
\begin{center}
\begin{tabular}{ccc}
phase boundary & location & unbroken $U(1)$\\
\hline
$(111)\leftrightarrow (011)$&$\text{Cone}_{\{2,3\}}$&$\{(g,g^2,g^3)\,|\,g\in U(1)\}$\\
$(011)\leftrightarrow (001)$&$\text{Cone}_{\{2,5\}}$&$\{(g,g^2,1)\,|\,g\in U(1)\}$\\
$(001)\leftrightarrow (000)$&$\text{Cone}_{\{4,5\}}$&$\{(g,1,1)\,|\,g\in U(1)\}$
\end{tabular}
\end{center}
Thus, to obtain the modules describing the lift defects for this type of path, our construction requires to introduce the following charge cutoffs in the module $\mathcal{T}_\infty$:
\begin{align}
\begin{split}
& Q^L_1+2Q^L_2+3Q^L_3\leq N \\
& Q^L_1+2Q^L_2\leq M \\
& Q^L_1\leq K.
\end{split}
\label{eq:cutoff(111)-(011)-(001)-(000)}
\end{align}
Let $\mathcal{T}_{N,M,K}^{(111)-(011)-(001)-(000)}$ be the 
submodule of $\mathcal{T}_\infty$ generated by all generators whose charges satisfy 
\eqref{eq:cutoff(111)-(011)-(001)-(000)}. Due to the relations
\begin{align}
\begin{split}
& X_2\alpha_1^{2}\alpha_2^{-1}=1 \\
& X_3\alpha_1^{-1}\alpha_2^{2}\alpha_3^{-1}=1\\
& X_4\alpha_2^{-1}\alpha_3^{2}=1
\end{split}
\label{eq:A3Tinftyrelations}
\end{align}
in $\mathcal{T}_\infty$ 
$\mathcal{T}_{N,M,K}^{(111)-(011)-(001)-(000)}$ is in fact finitely generated. Namely, the charges under the unbroken $U(1)$s at the phase boundaries of $\alpha_1^{2}\alpha_2^{-1}$, $\alpha_1^{-1}\alpha_2^{2}\alpha_3^{-1}$ and $\alpha_2^{-1}\alpha_3^{2}$ are given by 
\[
\begin{array}{c|ccc}
&Q^L_1+2Q^L_2+3Q^L_3&Q^L_1+2Q^L_2&Q^L_1\\
\hline
\alpha_1^{2}\alpha_2^{-1}&0&0&2\\
\alpha_1^{-1}\alpha_2^{2}\alpha_3^{-1}&0&3&-1\\
\alpha_2^{-1}\alpha_3^{2}&4&-2&0
\end{array}
\]
Hence, $\mathcal{T}_{N,M,K}^{(111)-(011)-(001)-(000)}$ is in fact generated by all the generators of $\mathcal{T}_\infty$ whose charges lie in the finite band
\begin{align}
\begin{split}
& N-4< Q^L_1+2Q^L_2+3Q^L_3\leq N \\
& M-3< Q^L_1+2Q^L_2\leq M \\
& K-2< Q^L_1\leq K
\end{split}
\end{align}
This matches with the band restriction rules for all homotopy classes of paths along $(111)\leftrightarrow (011)\leftrightarrow (001)\leftrightarrow (000)$, from \cite{herbst2008}. 

It is not difficult to describe $\mathcal{T}_{N,M,K}^{(111)-(011)-(001)-(000)}$ concretely.
For instance for $K\in 2\mathbb{Z}$ and $M\in 3\mathbb{Z}$ a set of independent generators is given by 
\begin{equation}
\alpha_1^K\alpha_2^{M-\frac{K}{2}}\alpha_3^{N-\frac{M}{3}}\{1,\alpha_1^{-1},\alpha_2^{-1},\alpha_3^{-1}\}.
\end{equation}

For the second type of paths $(111)\leftrightarrow (101)\leftrightarrow (001)\leftrightarrow (000)$ we proceed in the same fashion. The relevant unbroken gauge groups at the traversed phase boundaries in that case are given by
\begin{center}
\begin{tabular}{ccc}
phase boundary & location & unbroken gauge group\\
\hline
$(111)\leftrightarrow (101)$&$\text{Cone}_{\{2,4\}}$&$\{(g,g^2,ga)\,|\,g\in U(1),\,a\in\mathbb{Z}_2\}$\\
$(101)\leftrightarrow (001)$&$\text{Cone}_{\{1,2\}}$&$\{(1,1,g)\,|\,g\in U(1)\}$\\
$(001)\leftrightarrow (000)$&$\text{Cone}_{\{4,5\}}$&$\{(g,1,1)\,|\,g\in U(1)\}$
\end{tabular}
\end{center}
Thus, the truncation of $\mathcal{T}_\infty$ describing the lift defects for this type of path is given by
\begin{align}
\begin{split}
& Q^L_1+2Q^L_2+Q^L_3\leq N \\
& Q^L_3\leq M \\
& Q^L_1\leq K.
\end{split}
\label{eq:cutoff(111)-(101)-(001)-(000)}
\end{align}
We denote the submodule of $\mathcal{T}_\infty$ generated by generators whose charges satisfy the inequalities \eqref{eq:cutoff(111)-(101)-(001)-(000)} by $\mathcal{T}_{N,M,K}^{(111)-(101)-(001)-(000)}$.
Using the relations (\ref{eq:A3Tinftyrelations}) we again find that the latter module is generated by generators of charges in a finite charge band. Namely, the charges of $\alpha_1^{2}\alpha_2^{-1}$, $\alpha_1^{-1}\alpha_2^{2}\alpha_3^{-1}$ and $\alpha_2^{-1}\alpha_3^{2}$ under the unbroken $U(1)$s for this type of path are given by
\[
\begin{array}{c|ccc}
&Q^L_1+2Q^L_2+Q^L_3&Q^L_3&Q^L_1\\
\hline
\alpha_1^{2}\alpha_2^{-1}&0&0&2\\
\alpha_1^{-1}\alpha_2^{2}\alpha_3^{-1}&2&-1&-1\\
\alpha_2^{-1}\alpha_3^{2}&0&2&0
\end{array}
\]
and hence $\mathcal{T}_{N,M,K}^{(111)-(101)-(001)-(000)}$ is generated by the generators of $\mathcal{T}_\infty$ whose charges lie in the band
\begin{align}
\begin{split}
& N-2< Q^L_1+2Q^L_2+Q^L_3\leq N \\
& M-2< Q^L_3\leq M \\
& K-2< Q^L_1\leq K.
\end{split}
\end{align}
Again, this reproduces all band restriction rules for these type of paths from \cite{herbst2008}.

Also in this case the modules can be concretely described.
For instance for $M+K\in 2\mathbb{Z}$ a set of independent generators is given by 
\begin{equation}
\alpha_1^K\alpha_2^{N-\frac{M+K}{2}}\alpha_3^{M}\{1,\alpha_1^{-1},\alpha_3^{-1},\alpha_1^{-1}\alpha_2^{-1}\alpha_3\}.
\end{equation}

\subsubsection*{General $N$}

Since the phase structure of the $A_{N-1}$ model becomes increasingly more complicated for higher values of $N$ we will restrict our discussion of orbifold lifts to those associated to classes of paths 
$(00\dots 000)\leftrightarrow (00\dots 001)\leftrightarrow (00\dots 011)\leftrightarrow\dots\leftrightarrow (01\dots 111)\leftrightarrow (11\dots 111)$.

The unbroken gauge group at the $i$-th phase boundary traversed by this path is given by
\begin{equation}
\{(g,g^2,g^3,\dots,g^i,1,1,\dots,1)\vert g\in U(1)\}.
\end{equation}
By our construction, the respective lift defects are associated to the truncated submodule $\mathcal{T}_{M_1,\ldots,M_{N-1}}$ which is generated by those generators in $\mathcal{T}_\infty$, whose charges satisfy the inequalities
\begin{align}
\begin{split}
& Q^L_1+2Q^L_2+3Q^L_3+\dots+(N-3)Q^L_{N-3}+(N-2)Q^L_{N-2}+(N-1)Q^L_{N-1}\leq M_{N-1}\\
& Q^L_1+2Q^L_2+3Q^L_3+\dots+(N-3)Q^L_{N-3}+(N-2)Q^L_{N-2}\leq M_{N-2}\\
& Q^L_1+2Q^L_2+3Q^L_3+\dots+(N-3)Q^L_{N-3}\leq M_{N-3}\\
& \vdots\\
& Q^L_1+2Q^L_2\leq M_2\\
& Q^L_1\leq M_1
\end{split}
\end{align}
As in the $A_2$ and $A_3$ cases, we can use the relations 
\begin{align}
\begin{split}
& X_2\alpha_1^{2}\alpha_2^{-1}=1 \\
& X_3\alpha_1^{-1}\alpha_2^{2}\alpha_3^{-1}=1\\
& X_4\alpha_2^{-1}\alpha_3^{2}\alpha_4^{-1}=1\\
&\vdots\\
&X_{N-1}\alpha_{N-3}^{-1}\alpha_{N-2}^2\alpha_{N-1}^{-1}=1\\
&X_{N}\alpha_{N-2}^{-1}\alpha_{N-1}^2=1
\end{split}
\label{eq:AN-1Tinftyrelations}
\end{align}
in $\mathcal{T}_\infty$
to show that the generators of $\mathcal{T}_{M_1,\ldots,M_{N-1}}$ have charges in a finite charge band.
For this, we observe that the charges of the relevant monomials in the $\alpha_i$ are given by
\[
\begin{array}{c|ccccc}
&Q^L_1&Q^L_1+2Q^L_2&Q^L_1+2Q^L_2+3Q_3^L&\ldots&\sum_{i=1}^{N-1}iQ_i^L\\
\hline
\alpha_1^{2}\alpha_2^{-1}&2&0&\ldots&\ldots&0\\
\alpha_1^{-1}\alpha_2^{2}\alpha_3^{-1}&-1&3&0&\ldots&\vdots\\
\alpha_2^{-1}\alpha_3^{2}\alpha_4^{-1}&0&-2&4&&\vdots\\
\vdots&\vdots&\ddots&\ddots&\ddots&\vdots\\
\alpha_{N-2}^{-1}\alpha_{N-1}^2&0&\ldots&0&-(N-2)&N
\end{array}
\]
Therefore, $\mathcal{T}_{M_1,\ldots,M_{N-1}}$ is generated by generators of $\mathcal{T}_\infty$ whose charges lie in the band
\begin{align}
\begin{split}
& M_{N-1} - (N-1) \leq Q^L_1+2Q^L_2+\dots+(N-3)Q^L_{N-3}+(N-2)Q^L_{N-2}+(N-1)Q^L_{N-1}\leq M_{N-1}\\
& M_{N-2} - (N-2) \leq Q^L_1+2Q^L_2+\dots+(N-3)Q^L_{N-3}+(N-2)Q^L_{N-2}\leq M_{N-2}\\
& M_{N-3} - (N-3) \leq Q^L_1+2Q^L_2+\dots+(N-3)Q^L_{N-3}\leq M_{N-3}\\
& \vdots\\
& M_2 - 2 \leq Q^L_1+2Q^L_2\leq M_2\\
& M_1 - 1 \leq Q^L_1\leq M_1,
\end{split}
\label{eq:AN-1chargerestrictions}
\end{align}
Note that here, for convenience in solving the inequalities, we slightly changed the presentation of the charge bands as compared to the treatment of the $A_2$ and $A_3$ cases above, in that we 
added one to the lower bound and replaced the strict inequality with a $\leq$. 

The inequalities (\ref{eq:AN-1chargerestrictions}) can be solved iteratively from bottom to top plugging in the solutions of the previous inequalities into the next. By induction we find that there are $N$ solutions $(Q_1^L,Q_2^L\dots,Q_{N-1}^L)$ to (\ref{eq:AN-1chargerestrictions}) and that $\sum_{j=1}^{n}jQ_j^L$ assumes $n+1$ consecutive integer values on the solutions for all $1\leq n<N$. 
Starting from the last line we have that 
\begin{equation}
Q_1^L=M_1-1,M_1.
\end{equation}
Thus $Q_1^L$ assumes two consecutive integer values on all solutions of the inequalities. 
The solutions for $Q_1^L$ can then be plugged into the second to last line, which yields
\begin{equation}
\frac{M_2-Q_1^L-2}{2}\leq Q_2^L\leq \frac{M_2-Q_1^L}{2}.
\end{equation}
Thus, if $2$ does not divide $(M_2-Q_1^L)$ we have that $Q_2^L$ has only one solution namely
\begin{equation}
Q_2^L=\frac{M_2-Q_1^L-1}{2}.
\end{equation}
However, since $Q_1^L$ runs through $2$ consecutive integers, for one of them $2$ does divide $M_2-Q_1^L$. For this $Q_1^L$ there are two solutions for $Q_2^L$:
\begin{equation}
Q_2^L=\frac{M_2-Q_1^L-2}{2},\frac{M_2-Q_1^L}{2}.
\end{equation} 
This yields $3$ solutions, on which $Q_1^L+2Q_2^L$ assumes $3$ consecutive integer values.
We can now go on inductively. For $Q_n^L$ we have the inequality
\begin{equation}
\frac{M_n-n-\sum_{j=1}^{n-1}jQ_j^L}{n}\leq Q_n^L\leq \frac{M_n-\sum_{j=1}^{n-1}jQ_j^L}{n}.
\end{equation}
Thus, if $n$ does not divide $M_n-\sum_{j=1}^{n-1}jQ_j^L$ we find that $Q_n^L$ has only one solution.
However $\sum_{j=1}^{n-1}jQ_j^L$ assumes $n$ consecutive integers. So for one of the solutions, $M_n-\sum_{j=1}^{n-1}jQ_j^L$ is divisible by $n$, in which case there are two solutions. So the last $n$ inequalities have $n+1$ solutions, on which $\sum_{j=1}^{n}jQ_j^L$ take $n+1$ consecutive integer values. Therefore all the inequalities have $N$ solutions. 

For concreteness, let us pick $M_i=i$. Then the solutions to \eqref{eq:AN-1chargerestrictions} are given by
\begin{equation}
(Q_1^L,Q_2^L\dots,Q_{N-1}^L)\in\left\{(0,\ldots,0),(1,0,\ldots,0),(0,1,0,\ldots,0),\ldots,(0,\ldots,0,1)
\right\},
\end{equation}
i.e. either all the $Q_i^L$ are $0$, or one of them is $1$ and all the others are $0$. 
The corresponding generators are 
\begin{equation}
\{\alpha_0\coloneqq 1,\alpha_1,\alpha_2,\alpha_3,\dots,\alpha_{N-1}\},
\end{equation}
and they satisfy relations
\begin{equation}
X'_1\alpha_n=
\begin{cases}
X_1X_2\dots X_{n}\alpha_{n-1}, & \quad n=2,3,\dots,N\\
X_1\alpha_N, & \quad n=1
\end{cases}
\end{equation}
and
\begin{equation}
X'_{N+1}\alpha_n=
\begin{cases}
X_{n+2}X_{n+3}\dots X_{N+1}\alpha_{n+1}, & \quad n=1,2,\dots,N-1\\
X_2X_3\dots X_{N+1}\alpha_N, & \quad n=N.
\end{cases}
\end{equation}
Also for this model we find agreement with \cite{herbst2008}: The charge bands \eqref{eq:AN-1chargerestrictions} of the lift defects match with the band restriction rules, and the fusion of fractional branes $\bC[X_1,X_{N+1}]/\langle X_1,X_{N+1}\rangle\{[n]_N\}$ with $\mathcal{T}_{1,2,\ldots,N-1}$ agrees with the lifts of the respective D-branes derived in 8.4.2. in \cite{herbst2008}.
\section{GLSM Description of Flows Between Minimal Models}\label{sec:minmods}

$N=2$ superconformal minimal models ${\mathcal M}_k$, $k\in\mathbb{N}_0$ can be obtained as IR fixed points of Landau-Ginzburg models with a single chiral field $X$ and homogeneous superpotential $W=X^d$ for $d\in\mathbb{Z}_{\geq 2}$. The level of the respective minimal model is given by $k=d-2$. 

Minimal models exhibit relevant perturbations, which are captured by deformations of the superpotential of the respective Landau-Ginzburg model by lower degree monomials in $X$ 
\begin{equation}
W=X^d+\lambda_1 X^{d-1}+\lambda_2 X^{d-2}+...+\lambda_{d-2}X^{2}.
\end{equation} 
This implies that  all minimal models ${\mathcal M}_k$ can be obtained as IR fixed points of relevant flows starting in minimal models ${\mathcal M}_{k_0}$ at levels $k_0>k$.
Along the flows, supersymmetric vacua become massive and decouple together with the A-type supersymmetric branes carrying the respective charges \cite{Hori:2000ck, Maldacena:2001ky,Brunner:2007qu}. What happens to A-branes under the flows can be explicitly studied using a geometric realization in terms of Lefshetz thimbles ending in critical points of the superpotential. 

The mirror dual of a minimal model ${\mathcal M}_k$ is the $\mathbb{Z}_{d}$-orbifold of ${\mathcal M}_k$. Hence, mirror duals of minimal models can be obtained as IR fixed points of $\mathbb{Z}_d$-orbifolds of Landau-Ginzburg models with a single chiral field $X$ and superpotential $W=X^d$. Here, the orbifold group acts on $X$ by phase multiplication. 

Mirror symmetry interchanges A-type and B-type branes, so in the mirror model B-type branes decouple. In \cite{Brunner:2007ur} B-type defects encoding the flows between Landau-Ginzburg orbifolds were constructed, which via fusion in particular describe the behavior of B-branes under the flows. 

In this section, we will apply the construction of this paper to flows between Landau-Ginzburg orbifolds, and in particular re-derive the flow defects of \cite{Brunner:2007ur}. This is possible, 
because the mirror duals of the flows between minimal models can be modelled in a GLSM. For single flow lines this was explained in \cite{Brunner:2021cga}, but indeed the entire parameter spaces of the mirror dual of ${\mathcal M}_{k=d-2}$ can be described in a single GLSM. 
This model has $(d-1)$ chiral fields $X_0,\ldots,X_{d-2}$ and superpotential $W=\prod_{i=0}^{d-2}X_i^{d-i}$.\footnote{Indeed, one can also consider the case with zero superpotential. The respective GLSM describes flows between $\mathbb{C}/\mathbb{Z}_d$-orbifold models.} The gauge group is $U(1)^{d-2}$, and the respective $(d-1)$ FI parameters correspond to the coupling constants in the minimal model orbifolds. The chiral fields are charged under the gauge group as follows:
\begin{equation}\label{eq:fullmodel}
\begin{array}{c|cccccccc}
&X_0&X_1&X_2&X_3&\ldots&\ldots&X_{d-3}&X_{d-2}\\
\hline
U(1)_0&(d-1)&-d&0&\ldots&\ldots&\ldots&\ldots&0\\
U(1)_1&1&-2&1&0&\ldots&\ldots&\ldots&0\\
U(1)_2&0&1&-2&1&0&\ldots&\ldots&0\\
U(1)_3&0&0&1&-2&1&0&\ldots&0\\
\vdots&\vdots&&\ddots&\ddots&\ddots&\ddots&\ddots&\vdots\\
U(1)_{d-4}&0&\ldots&\ldots&0&1&-2&1&0\\
U(1)_{d-3}&0&\ldots&\ldots&\ldots&0&1&-2&1\\
\end{array}
\end{equation}
Note that only $U(1)_0$ is anomalous, whereas the other $U(1)_i$, $i>0$ are non-anomalous. Hence only the FI parameter of $U(1)_0$ has a non-trivial RG flow, whereas the other ones are honest parameters of the theory.\footnote{
Indeed, there is a  similarity between this GLSM and the $A_{d-2}$ model describing the non-linear sigma model on the resolution of the $A_{d-2}$ singularity, c.f.~Section~\ref{sec:anmodel}. This is not unexpected as the LG model is taylored to describe the physics of the singularity, and our model is in turn related to that one by mirror symmetry.
Apart from the existence of the anomalous $U(1)_0$ in the minimal model GLSM, both models 
are basically identical. They have the same gauge group and the same number of chiral fields of the same charges. Besides the anomalous $U(1)_0$ the only other difference is the non-trivial superpotential in the minimal model GLSM. Setting this to zero leads to a GLSM describing the parameter space of the $\mathbb{C}/\mathbb{Z}_d$ orbifold model, which has very similar properties to the minimal model GLSM. It would be very interesting to understand the relationship between the minimal model GLSM and the $A_{d-2}$ model better.} 

This GLSM exhibits $(d-1)$ phases. In each of these phases all but one of the chiral fields  assume a non-trivial vacuum expectation value. We call the phase in which only the field $X_i$ does not assume a vev phase $i$. In the effective field theory describing this phase, only that field remains, the superpotential becomes $W=X_{i}^{d-i}$ and the gauge group is broken to $\mathbb{Z}_{d-i}$. Hence in phase $i$ the GLSM is effectively described by the Landau-Ginzburg orbifold whose IR fixed point is the mirror dual of the minimal model ${\mathcal M}_k$ at level $k=d-i-2$. Thus, the GLSM contains as phases all the minimal model orbifolds of levels up to $k=d-2$. 

Every such phase is separated from any other by a codimension-one phase boundary. The one separating phases $i$ and $j$ are located at the cones $\text{Cone}_{\{k\notin\{i,j\}\}}$ consisting of the positive linear combinations of the charge vectors of all chiral fields $X_k$, $k\notin\{i,j\}$.

Having realized the entire parameter spaces of minimal models in an abelian GLSM, we can now apply the strategy outlined in Section~\ref{sec:Review} to construct defects lifting the Landau-Ginzburg orbifolds describing the phases of the GLSM to the GLSM. These can in particular be used to obtain the flow defects between different minimal model orbifolds. 
The procedure is exactly the same as in the non-anomalous case: We start out from the GLSM identity defect, then push the model down to a phase by setting fields to their vevs on one side of the defect. 
We introduce charge truncations for every phase boundary to be traversed by a chosen path, which then gives rise to a charge band. While the procedure is completely the same in anomalous and non-anomalous cases, there is a notable difference in the outcome however: In contrast to the non-anomalous case, one obtains different charge bands from lifting the UV and the IR phases of a given flow. Indeed, lifting IR phases yields bands which are strictly smaller than the ones from the respective UV lifts. This is in qualitative agreement with the discussion of D-brane transport in other anomalous GLSMs in \cite{Hori:2013ika,Clingempeel:2018iub}, where it was found that the D-brane transport is goverend by a `large' and a `small' charge window. D-branes in the large window can be transported along the flow, but the ones which are not in the small window undergo some kind of decay. 

Before giving the general construction, we will first spell out the details in a ``truncated" example. RG flows of minimal models come in a hierarchy. The respective perturbations can be restricted to the $i$ least relevant ones for $1\leq i<d-1$. On the level of Landau-Ginzburg models this corresponds to restricting the deformation of the superpotential to $W=X^d+\lambda_1 X^{d-1}+\ldots+\lambda_i X^{d-i}$. The respective flow drives the minimal model at level $k=d-2$ at most to the one at level $k=d-2-i$. We call this the $i$-step perturbations. Indeed, one can easily obtain GLSMs capturing these restricted parameter spaces by freezing the fields $X_j$, $j>i$ to vevs in the GLSMs describing the general perturbations. This procedure breaks the gauge groups $U(1)_j$ for $j>i$. 
The resulting GLSM has chiral fields $X_0, \ldots, X_i$, superpotential $W=X_0^dX_1^{d-1}\ldots X_{i}^{d-i}$ and gauge group $U(1)^i$. The charges of the fields $X_i$ under the gauge group are given by 
$$
\begin{array}{c|cccccccc}
&X_0&X_1&X_2&X_3&\ldots&\ldots&X_{i-1}&X_{i}\\
\hline
U(1)_0&(d-1)&-d&0&\ldots&\ldots&\ldots&\ldots&0\\
U(1)_1&1&-2&1&0&\ldots&\ldots&\ldots&0\\
U(1)_2&0&1&-2&1&0&\ldots&\ldots&0\\
U(1)_3&0&0&1&-2&1&0&\ldots&0\\
\vdots&\vdots&&\ddots&\ddots&\ddots&\ddots&\ddots&\vdots\\
U(1)_{i-1}&0&\ldots&\ldots&0&1&-2&1&0\\
U(1)_{i}&0&\ldots&\ldots&\ldots&0&1&-2&1\\
\end{array}
$$
Below we will give the concrete construction of the defects embedding the phases into the GLSM for  two-step perturbations and derive the respective flow defects between the minimal models.

Note that one can indeed construct GLSMs describing other subspaces of the minimal model parameter space by freezing arbitrary combintations of fields to vevs in the GLSM containing the entire parameter space. Freezing for instance all but the fields $X_0$ and $X_i$, one arrives at a $U(1)$-GLSM describing a specific one-parameter flow from the the minimal model of level $k=d-2$ to the one at level $k=d-2-i$.\footnote{These are the models studied in \cite{Brunner:2021cga}.}

\subsection{GLSM for the Two-Step Minimal Model Flows}

As discussed above, the two-step flows starting in the minimal model of level $k=d-2$ can be modelled in the gauged linear sigma model with gauge group $U(1)^2$, chiral fields $X_0,X_1,X_2$ of gauge charges
$$
\begin{array}{c|ccc}
&X_0&X_1&X_2\\
\hline
U(1)_0&(d-1)&-d&0\\
U(1)_1&1&-2&1
\end{array},
$$
and superpotential $W=X_0^dX_1^{d-1}X_2^{d-2}$. 
This model exhibits three Landau-Ginzburg orbifold phases. In each of those two of the three chiral fields assume a non-trivial vaccum expectation value, and only one field remains part of the effective low energy theory. The gauge group is broken to a finite subgroup. The phases are respectively described by the $\mathbb{Z}_{d-i}$-orbifolds of the Landau-Ginzburg theories with one chiral field $X_i$ and superpotential $W=X_i^{d-i}$, where $i\in\{0,1,2\}$. The orbifold group acts by phase multiplication on the field $X_i$. 

The phase boundary between phases $i$ and $j$ is located at the ray $\text{Cone}_{\{k\}}=\mathbb{R}^{\geq 0} Q_k$ in the direction of the charge of the chiral field $X_k$, $k\notin\{i,j\}$ which assumes a non-trivial vev in both phases $i$ and $j$.  The phase diagram is given by

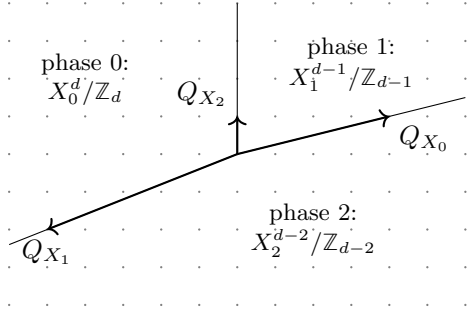
\begin{figure}[h]
\center
\begin{tikzpicture}
\foreach \x in {-6,-5,...,6} {
\foreach \y in {-4,-3,...,4} {
\fill[color=gray] (\x/2,\y/2) circle (0.15mm);
}
}
\draw[thick,->] (0,0) -- (2,0.5) node[anchor=north west] {$Q_{X_0}$};
\draw[thick,->] (0,0) -- (-5/2,-1) node[anchor=north] {$Q_{X_1}$};
\draw[thick,->] (0,0) -- (0,1/2) node[anchor=south east] {$Q_{X_2}$};
\draw[thick] (1.5,1.2) node[] {\small \begin{tabular}{c}phase $1$:\\ $X_1^{d-1}/\mathbb{Z}_{d-1}$\end{tabular}};
\draw[thick] (1,-1) node[] {\small \begin{tabular}{c}phase $2$:\\ $X_2^{d-2}/\mathbb{Z}_{d-2}$\end{tabular}};
\draw[thick] (-2,1) node[] {\small \begin{tabular}{c}phase $0$:\\ $X_0^d/\mathbb{Z}_d$\end{tabular}};
\draw[thin,-] (0,0) -- (0,2);
\draw[thin,-] (0,0) -- (6/2,1.5/2) ;
\draw[thin,-] (0,0) -- (-6/2,-2.4/2);
\end{tikzpicture}
\caption{Phase diagram of the GLSM describing the two-step minimal model flows.}
\label{mmPhases}
\end{figure}
The preserved $U(1)$-gauge groups on the phase boundaries are the stabilizers of the respective fields having non-trivial vev in both the adjacent phases:

\begin{equation}
\begin{array}{c|c}
\text{Phase boundary} & \text{preserved gauge group}\\
\hline
(01)&\{(g,1)|g\in U(1)\}=U(1)_0\\
(12)&\{(g,g^{-(d-1)})|g\in U(1)\}\\
(02)&\{(g^2,g^{-d})|g\in U(1)\}
\end{array}
\end{equation}
 
\subsection{Lift Defects for the Two-Step Model}

In the following we will construct the defects lifting the  Landau-Ginzburg orbifold model $(W=X_0^d/\mathbb{Z}_d)$ of phase $0$ to the GLSM. 
Starting point is the GLSM identity defect. The respective module is given by
\begin{equation}
{\mathcal I}_\text{GLSM}={{S\otimes V}\over \langle (X_0-\alpha^{d-1}\beta X_0^\prime), (X_1-\alpha^{-d}\beta^{-2} X_1^\prime), (X_2-\beta X_2^\prime) \rangle}.
\end{equation}
Here 
\begin{eqnarray}
R&=&\mathbb{C}[X_0,X_1,X_2,X_0^\prime,X_1^\prime,X_2^\prime]\\
S&=&{R\over \langle X_0^dX_1^{d-1}X_2^{d-2}-(X_0^\prime)^d(X_1^\prime)^{d-1}(X_2^\prime)^{d-2}\rangle}\\
V&=&{\mathbb{C}[\alpha,\alpha^{-1},\beta,\beta^{-1}]\over \langle (\alpha\alpha^{-1}-1), (\beta\beta^{-1}-1)\rangle}.
\end{eqnarray}
The charges of the auxiliary fields $\alpha,\alpha^{-1},\beta,\beta^{-1}$ under the gauge groups of the left, respectively right of the defect are given by
\begin{equation}
\begin{array}{c|cccc}
&\alpha&\alpha^{-1}&\beta&\beta^{-1}\\
\hline
Q_0^L& 1 &-1&0&0\\
Q_0^R& -1&1&0&0\\
Q_1^L& 0 &0&1&-1\\
Q_1^R& 0&0&-1&1
\end{array}
\end{equation}
In phase $0$ the fields $X_1$ and $X_2$ both have non-trivial vacuum expectation value. Hence, to obtain the defect embedding this phase into the GLSM, we have to set $X_1^\prime=1=X_2^\prime$ in ${\mathcal I}_\text{GLSM}$. This yields the module
\begin{equation}
{\mathcal T}_\infty={{\widetilde{S}\otimes V}\over \langle (X_0-\alpha^{d-1}\beta X_0^\prime), (X_1-\alpha^{-d}\beta^{-2}), (X_2-\beta) \rangle}.
\end{equation}
with 
\begin{eqnarray}
\widetilde{R}&=&\mathbb{C}[X_0,X_1,X_2,X_0^\prime]\\
\widetilde{S}&=&{\widetilde{R}\over \langle X_0^dX_1^{d-1}X_2^{d-2}-(X_0^\prime)^d\rangle}.
\end{eqnarray}
Next we have to impose the cutoffs. These depend on cutoff parameters which correspond to 
the chosen homotopy class of paths in parameter space. We would like to construct embeddings valid for flows from the UV phase (phase $0$) all the way to the IR phase (phase $2$). But there are two possibilities to go from phase $0$ to phase $2$. Either one can pass phase $1$ on the way, or one can avoid it, crossing directly from phase $0$ to phase $2$. We will first discuss the case, in which phase $1$ is passed, hence two phase boundaries, $(01)$ and $(12)$ are crossed.
For each phase boundary, a cutoff is introduced. 

On the phase boundary $(01)$ $U(1)_0$ is preserved, implying a cutoff 
\begin{equation}\label{eq:bound01}
Q_{(01)}^L:=Q_0^L\leq N_{(01)}
\end{equation}
for a choice of $N_{(01)}\in\mathbb{Z}$. On the phase boundary $(12)$ the group $\{(g,g^{-(d-1)})|g\in U(1)\}\cong U(1)$ is preserved leading to the additional cutoff 
\begin{equation}\label{eq:bound12}
Q_{(12)}^L:=Q_0^L-(d-1)Q_1^L\leq N_{(12)}
\end{equation}
for a choice of $N_{(12)}\in\mathbb{Z}$. 

Introduction of the cutoffs means that one considers the submodule ${\mathcal T}_{N_{(01)},N_{(12)}}$ of ${\mathcal T}_\infty$ generated over the algebra $\widetilde{S}$ by only those generators whose charges satisfy the inequalities (\ref{eq:bound01}) and (\ref{eq:bound12}). Due to the relations
\begin{equation}\label{eq:tinftyrelations}
X_1\alpha^{d}\beta^{2}=1,\quad X_2\beta^{-1}=1\,.
\end{equation} 
holding in the module ${\mathcal T}_\infty$ all generators in the truncated submodule can be obtained by applying the algebra $\widetilde{S}$ on generators whose charges lie in the band
\begin{equation}\label{eq:band0112}
\begin{array}{rcccl}
N_{(01)}-d&<&Q_{(01)}^L:=Q_0^L&\leq& N_{(01)}\\
N_{(12)}-d+1&<&Q_{(12)}^L:=Q_0^L-(d-1)Q_1^L&\leq& N_{(12)}
\end{array}
\end{equation}
Namely, using the first relation in (\ref{eq:tinftyrelations}) any generator $e$ of ${\mathcal T}_{N_{(01)},N_{(12)}}$ of charges $Q_{(01)}^L(e)\leq N_{(01)}-d$ can be written as
\begin{equation}
e=(X_1\alpha^d\beta^2) e=:X_1 e^\prime
\end{equation}
where $e^\prime\in {\mathcal T}_{N_{(01)},N_{(12)}}$ has charges
\begin{equation}
Q_{(01)}^L(e^\prime)=Q_{(01)}^L(e)+d
\qquad
Q_{(12)}^L(e^\prime)=Q_{(12)}^L(e)-(d-2)
\end{equation}
Hence, successively, $e=X_1^m e^{\prime\prime}$ with $e^{\prime\prime}\in {\mathcal T}_{N_{(01)},N_{(12)}}$ and $N_{(01)}-d<Q_{(01)}^L(e^{\prime\prime})\leq N_{(01)}$. One can now use the second relation in (\ref{eq:tinftyrelations})  in a similar fashion to write
\begin{equation}
e^{\prime\prime}=(X_2\beta^{-1})^ne^{\prime\prime}=:X_2^n e^{\prime\prime\prime}
\end{equation}
for some $n\in\mathbb{N}_0$, such that 
\begin{equation}
N_{(01)}-d<Q_{(01)}^L(e^{\prime\prime\prime})\leq N_{(01)}\qquad
N_{(12)}-d+1<Q_{(12)}^L(e^{\prime\prime\prime})\leq N_{(12)}
\end{equation}
Hence, any generator of ${\mathcal T}_{N_{(01)},N_{(12)}}$ can be obtained by applying an element of $\widetilde{S}$ on a generator whose charges satisfy (\ref{eq:band0112}). Thus, the truncated module ${\mathcal T}_{N_{(01)},N_{(12)}}$
is generated by generators in this charge band.

In the following we will describe this submodule in more detail. It depends on the choice of the two integers $N_{(01)}$ and $N_{(12)}$. Define $z\in\mathbb{Z}$ and $n\in\{1,\ldots d-1\}$ such that 
\begin{equation}\label{eq:Nparam}
N_{(01)}-N_{(12)}=z(d-1)+n.
\end{equation}
The first line of (\ref{eq:band0112}) implies that the charge $Q_0\in\{1,\ldots d-1\}$ of generators is given by
\begin{equation}
Q_0\in\{1,\ldots d-1\}=N_{(01)}-i,\quad i\in\{0,\ldots d-1\}.
\end{equation}
Substituting this into the second line in (\ref{eq:band0112}) and using the parametrization (\ref{eq:Nparam}) one obtains
\begin{equation}
(z+1-Q_1^L)(d-1)+n>i\geq (z-Q_1^L)(d-1)+n.
\end{equation}
The solutions to the inequalities (\ref{eq:band0112}) can now be read off as
\begin{equation}
i\in\{0,1,\ldots, n-1\}\quad\text{for}\quad Q_1^L=z+1\qquad\text{and}\qquad
i\in\{n,n+1,\ldots, d-1\}\quad\text{for}\quad Q_1^L=z\,.
\end{equation}
The truncated module ${\mathcal T}_{N_{(01)},N_{(12)}}$ is therefore generated by 
\begin{equation}
\alpha^{N_{(01)}}\beta^z\left\{\beta,\beta\alpha^{-1},\beta\alpha^{-2},\ldots,\beta\alpha^{-(n-1)},\alpha^{-n},\alpha^{-n-1},\ldots,\alpha^{-(d-1)}\right\}
\end{equation}
Denote the generators by 
\begin{equation}
e_i:=\alpha^{N_{(01)}}\beta^{z}\alpha^{-i}
\cdot\left\{
\begin{array}{ll}\beta,&\text{if } 0\leq i<n\\
1,&\text{if } n\leq i<d
\end{array}
\right.
\end{equation}
Using the relations in the module ${\mathcal T}_\infty$ it is not difficult to find the relations in the truncated module ${\mathcal T}_{N_{(01)},N_{(12)}}$. For $i\in\{1,\ldots, n-1,n+1,\ldots,d-1\}$ one obtains
\begin{equation}
X_0^\prime e_i=X_0\alpha^{-(d-1)}\beta^{-1}  e_i=X_0X_1 \alpha\beta e_i=X_0X_1X_2\alpha e_i=X_0X_1X_2 e_{i-1}\,,
\end{equation}
whereas
\begin{equation}
\begin{array}{l}
X_0^\prime e_n=X_0\alpha^{-(d-1)}\beta^{-1} e_n=X_0X_1\alpha\beta e_n=X_0X_1 e_{n-1}\\
X_0^\prime e_0=X_0\alpha^{-(d-1)}\beta^{-1} e_0=X_0e_{d-1}\,.
\end{array}
\end{equation}
The module ${\mathcal T}_{N_{(01)},N_{(12)}}$ can be obtained as the cokernel of the matrix
\begin{equation}
p_1=\left(\small\begin{array}{ccccccccc}
X_0^\prime&&&&&&&& -X_0X_1X_2\\
-X_0&X_0^\prime &&&&&&&\\
&-X_0X_1X_2&\ddots &&&&&&\\
&&\ddots&X_0^\prime&&&&&\\
&&&-X_0X_1X_2&X_0^\prime&&&&\\
&&&&-X_0X_1&X_0^\prime&&&\\
&&&&&-X_0X_1X_2&\ddots &&\\
&&&&&&\ddots&\ddots&\\
&&&&&&&-X_0X_1X_2&X_0^\prime
\end{array}
\right)
\end{equation}
on a free module of rank $d$ whose generators have the same charges as the $e_i$. Note that in this matrix the elements on the secondary diagonal are $-X_0X_1X_2$ except in two places in one of which it is $-X_0$ and in the other one $-X_0X_1$. The positions of these two entries is determined by the cutoff parameters $N_{(01)}$ and $N_{(12)}$.

This module corresponds to the defect lifting phase $0$ into the GLSM in a way compatible with flows from phase $0$ via phase $1$ to phase $2$. In fact, one can find a suitable $d\times d$-matrix $p_0$ which completes $p_1$ to a $U(1)^d\times\mathbb{Z}_d$-equivariant matrix factorization of $X_0^dX_1^{d-1}X_2^{d-2}-(X_0^\prime)^d$.

Indeed, these lift defects are consistent with what is known about flows between minimal models. Pushing the GLSM to phase $2$ on the left of the defect should reproduce the flow defects between minimal models, which have been constructed in \cite{Brunner:2007ur}. Indeed, this is accomplished by just setting $X_0=1=X_1$ in the truncated module ${\mathcal T}_{N_{(01)},N_{(12)}}$. In this way one obtains the cokernel of the matrix
\begin{equation}\label{eq:rgmatrix}
\left(\small\begin{array}{ccccccccc}
X_0^\prime&&&&&&&& -X_2\\
-1&X_0^\prime &&&&&&&\\
&-X_2&\ddots &&&&&&\\
&&\ddots&X_0^\prime&&&&&\\
&&&-X_2&X_0^\prime&&&&\\
&&&&-1&X_0^\prime&&&\\
&&&&&-X_2&\ddots &&\\
&&&&&&\ddots&\ddots&\\
&&&&&&&-X_2&X_0^\prime
\end{array}
\right)
\end{equation}
Note that this matrix contains two entries $-1$ on the secondary diagonal. These can be used to reduce the presentation of the respective module. It can be written as a cokernel of a $(d-2)\times(d-2)$-matrix whose only non-zero entries are on the diagonal  and the secondary diagonal. The entries on the diagonal are $X_2$ and the ones on the secondary diagonal are $-X_0$ except that depending on the choices of cutoff parameters either in two places we have $-X_0^2$ instead of $-X_0$, or in one place $-X_0^3$ instead of $-X_0$. These are exactly the defects describing the flows from the minimal model orbifold at level $k=d-2$ to the one at level $k=d-4$ \cite{Brunner:2007ur}.
Thus, the lift defects provide all the flow defects from phase $0$ to phase $2$.

As alluded to above, one can also cross over directly from phase $0$ to phase $2$ in the GLSM. Before constructing the lift defects associated to such paths, let us briefly pause and discuss the action of the lift defects on D-branes. To illustrate it, let us determine the lifts of the elementary D-branes associated to the $S':=\mathbb{C}[X'_0]/\langle (X_0')^d \rangle$-modules
\[
{\mathcal M}_m:=S'/X'_0 S' \{ [m]_d \}\,,\quad m=0,\ldots d-1.
\] 
The lifts are given by the $\mathbb{Z}_d$-invariant parts of the tensor products ${\mathcal T}_{N_{(01)},N_{(12)}}\otimes_{S'}\mathcal{M}_m$. They are straight-forward to calculate. The $\mathbb{Z}_d$-projection singles out a single generator for each $m$. Let $i:=N_{(01)}-m\,\text{mod}\,d\in\{0,\ldots, d-1\}$. Then, one obtains
the $S''=\mathbb{C}[X_0,X_1,X_2]/\langle X_0^dX_1^{d-1}X_2^{d-2}\rangle$-module
\begin{equation}\label{eq:windowproj}
\begin{array}{ll}
S''/X_0X_1X_2S''\{(N_{(01)}-i,z+1)\}&\text{if}\;0\leq i<n-1\\
S''/X_0X_1S''\{(N_{(01)}-i,z+1)\}&\text{if}\; i=n-1\\
S''/X_0X_1X_2S''\{(N_{(01)}-i,z)\}&\text{if}\;n\leq i<d-1\\
S''/X_0S''\{(N_{(01)}-i,z)\}&\text{if}\;i=d-1\,.
\end{array}
\end{equation}
Under the flow to the IR phase, which is implemented by setting $X_0=1=X_1$ $d-2$ of the elementary D-branes in the UV phase are mapped to elementary D-branes in the IR, and two $(i=n-1,d-1)$ are mapped to trivial D-branes. The respective D-branes decouple.\footnote{All the lifts of UV D-branes in \eqref{eq:windowproj} lie in the large window, but only the ones with $i\notin\{n-1,d-1\}$ lie in the small window.}

After this aside about D-branes, let us return to the construction of lift defects. The lift defects associated to paths directly crossing from phase $0$ to phase $2$ are obtained 
 from ${\mathcal T}_\infty$ by only a single truncation in the direction of the $U(1)$ preserved along the phase boundary $(02)$. 
The latter is given by $\{(g^2,g^{-d})|g\in U(1)\}\cong U(1)$ leading to the truncation 
\begin{equation}\label{eq:bound02}
Q_{(02)}^L:=2Q_0^L-dQ_1^L\leq N_{(02)}.
\end{equation}
As before, via the relations (\ref{eq:tinftyrelations}), all generators in the truncated module are obtained by action of $\widetilde{S}$ on generators whose charges lie in the band
\begin{equation}\label{eq:band02}
N_{(02)}-d<Q_{(02)}^L:=2Q_0^L-dQ_1^L\leq N_{(02)}
\end{equation}
Since this is only a restriction in one direction on a rank-$2$ lattice of charges, there are still infinitely many generators. Writing
\begin{equation}
N_{(02)}=zd+n\quad\text{for}\;z\in\mathbb{Z}\;\text{and}\;n\in\{0,\ldots,d-1\},
\end{equation}
one can easily find the set of charges $(Q_0^L,Q_1^L)$ satisfying 
 (\ref{eq:band02}):
 \begin{equation}
\begin{array}{rl}
&\{(-i-md,-2m-z)\,|\,0\leq i<\frac{d-n}{2},\,m\in\mathbb{Z}\}\\
\cup&
\{(-i-md,-2m-z-1)\,|\,\frac{d-n}{2}\leq i<\frac{2d-n}{2},\,m\in\mathbb{Z}\}\\
\cup&
\{(-i-md,-2m-z-2)\,|\,\frac{2d-n}{2}\leq i<d,\,m\in\mathbb{Z}\}
\end{array}
\end{equation}
Thus, the generators of the truncated module ${\mathcal T}_{N_{(02)}}$ are given by $e_{i,m}$ for $i\in\{0,\ldots, d-1\}$ and $m\in\mathbb{Z}$ with
\begin{equation}
e_{i,m}:=\beta^{-z}(\alpha^{d}\beta^{2})^{-m}\alpha^{-i}\cdot\left\{
\begin{array}{ll}
1,&\text{if}\;0\leq i<\frac{d-n}{2}\\
\beta^{-1},&\text{if}\;\frac{d-n}{2}\leq i<\frac{2d-n}{2}\\
\beta^{-2},&\text{if}\;\frac{2d-n}{2}\leq i<d
\end{array}
\right.
\end{equation}
The relations in ${\mathcal T}_{N_{(02)}}$ can be written as\footnote{$\lceil r\rceil$ denotes the ceiling of $r$, i.e.~the smalles integer $\geq r$.}
\begin{equation}
X_1e_{i,m}=e_{i,m+1}\,,\;\;\text{and}\;\;
X_0^\prime e_{i,m}=\left\{\begin{array}{ll} X_0X_1X_2 \,e_{i-1,m}&\text{if}\;i\notin\{0,\lceil \frac{d-n}{2}\rceil,\lceil\frac{2d-n}{2}\rceil\}\\
X_0X_1 \,e_{i-1,m}&\text{if}\;i\in\{\lceil \frac{d-n}{2}\rceil,\lceil\frac{2d-n}{2}\rceil\}\\
X_0X_2 \,e_{i-1,m}&\text{if}\;i=0\\
\end{array}\right.
\end{equation}
In contrast to the lift compatible with the flow from phase $0$ via phase $1$ to phase $2$, this module corresponds to a matrix factorization of infinite rank. Pushing to phase $2$ on the left side of the corresponding defect amounts to setting $X_0=1=X_1$. This renders the rank of the module (resp.~the matrix factorization) finite. Indeed, as for the case of flows passing phase $1$, the resulting module can be written as a cokernel of a matrix of the form (\ref{eq:rgmatrix}), which however is not as general as in the case of the flows passing through phase $1$. Namely, the spacing of the positions of the scalars $-1$ on the secondary diagonal is fixed by $d$, whereas in the case of the flow passing through phase $1$ it depends on the choice of cutoff parameter, and thereby on the chosen path in the parameter space. Thus, one only gets some flow defects between phases $0$ and $2$. 

Indeed, there is another GLSM describing the direct flow between phases $0$ and $2$--the GLSM obtained from the two-step model by giving $X_1$ a vaccum expectation value. This GLSM has only the fields $X_0,X_2$. Its gauge group is the stabilizer of $X_1$, and hence exactly the $U(1)$ in the two-step model which is preserved on the phase boundary between phases $0$ and $2$. The model has two phases, corresponding to phase $0$ and $2$ of the two-step model. We can now similarly lift phase $0$ to this model. This requires the choice of a single cutoff. 
The resulting defects are of finite rank, and easy to construct. Fusing these lift defects  with the defect between the GLSM with only fields $X_0,X_2$ and the two-step GLSM obtained by setting $X_1=1$ in the identity defect of the two-step GLSM, we obtain exactly the lift defects in the two-step model which are compatible with the direct crossover from phase $0$ to phase $2$. 

\subsection{Lift Defects for the Full Model}

In an analogous fashion  one can construct lift defects for the GLSM describing the full parameter space of minimal model orbifolds. We will briefly discuss the construction of defects lifting the UV phase $X^d/\mathbb{Z}_d$ into the full model, the field content of which is given in equation  (\ref{eq:fullmodel}). 
Our construction works for all possible paths in parameter space, but we will restrict the discussion to paths crossing all the phase boundaries $(i,i+1)$, $i=0,\ldots,d-3$. Those are paths which traverse all possible phases starting in the UV phase $0$ and then step by step passing phases $1$, $2$, etc. going all the way down to the trivial phase $(d-2)$. 

As before, the construction starts with the GLSM identity defect. The associated module is given by
\begin{equation}
{\mathcal I}_\text{GLSM}={{S\otimes V}\over 
\left\langle\begin{array}{c} (X_0-\alpha_0^{d-1}\alpha_1 X_0^\prime), (X_1-\alpha_0^{-d}\alpha_1^{-2}\alpha_2 X_1^\prime),\\
 (X_2-\alpha_1\alpha_2^{-2}\alpha_3 X_2^\prime),\ldots,(X_{d-4}-\alpha_{d-5}\alpha_{d-4}^{-2}\alpha_{d-3}X_{d-4}^\prime),\\
  (X_{d-3}-\alpha_{d-4}\alpha_{d-3}^{-2}X_{d-3}^\prime),(X_{d-2}-\alpha_{d-3}X_{d-2}^\prime)
  \end{array}\right\rangle}.
\end{equation}
Here 
\begin{eqnarray}
R&=&\mathbb{C}[X_0,\ldots,X_{d-2},X_0^\prime,\ldots,X_{d-2}^\prime]\\
S&=&{R\over \langle X_0^dX_1^{d-1}\cdot\ldots\cdot X_{d-2}^{2}-(X_0^\prime)^d(X_1^\prime)^{d-1}\cdot\ldots\cdot(X_{d-2}^\prime)^{2}\rangle}\\
V&=&{\mathbb{C}[\alpha_0,\alpha_0^{-1},\ldots,\alpha_{d-3},\alpha_{d-3}^{-1}]\over \langle (\alpha_0\alpha_0^{-1}-1),\ldots,(\alpha_{d-3}\alpha_{d-3}^{-1}-1) \rangle}
\end{eqnarray}
The $\alpha_i$ have charges $1$ and $-1$ with respect to $U(1)_i$ to the left, respectively right of the defect, and are not charged under the other $U(1)_j$, $j\neq i$.

Going down to phase $0$ on the right of the defect amounts to setting $X_i^\prime=1$ for $i\neq 0$ in this module. This yields
\begin{equation}\label{eq:tinftyfull}
{\mathcal T}_\infty={{\widetilde{S}\otimes V}\over
\left\langle\begin{array}{c} (X_0-\alpha_0^{d-1}\alpha_1 X_0^\prime), (X_1-\alpha_0^{-d}\alpha_1^{-2}\alpha_2 ),\\
 (X_2-\alpha_1\alpha_2^{-2}\alpha_3),\ldots,(X_{d-4}-\alpha_{d-5}\alpha_{d-4}^{-2}\alpha_{d-3}),\\
  (X_{d-3}-\alpha_{d-4}\alpha_{d-3}^{-2}),(X_{d-2}-\alpha_{d-3})
  \end{array}\right\rangle
}
\end{equation}
with 
\begin{eqnarray}
\widetilde{R}&=&\mathbb{C}[X_0,\ldots,X_{d-2},X_0^\prime]\\
\widetilde{S}&=&{\widetilde{R}\over \langle X_0^dX_1^{d-1}\cdot\ldots\cdot X_2^{2}-(X_0^\prime)^d\rangle}.
\end{eqnarray}
Next, we have to impose a cutoff in this module, for every phase boundary $(i,i+1)$ traversed by the chosen path. The $U(1)$ gauge group preserved on this phase boundary is just the stabilizer of all the chiral fields $X_j$, $j\notin\{i,i+1\}$ obtaining a vev in both phases $i$ and $(i+1)$.
It is given by
\begin{equation}
U(1)\cong\{(g,g^{-(d-1)},g^{-(d-2)},\ldots,g^{-(d-i)},1,\ldots,1)\,|\,g\in U(1)\}\subset U(1)^{d-2}.
\end{equation}
The respective cutoff reads
\begin{equation}
Q_{(i\,i+1)}^L:=Q_0^L-\sum_{j=1}^i(d-j)Q_j^L\leq N_{(i\,i+1)}
\end{equation}
for a choice of cutoff parameter $N_{(i\,i+1)}$. We denote the submodule generated by all generators of ${\mathcal T}_\infty$ satisfying these bounds by ${\mathcal T}_{N_{(0\,1)},\ldots,N_{(d-3\,d-2)}}$. 

By virtue of the relations for $X_i$, $i>0$ in (\ref{eq:tinftyfull}) the charges of generators of the truncated module actually lie in a band. The size of the band can be read off from the charges of the $X_i$, $i>1$:
\begin{equation}
\begin{array}{c|cccccc}
&Q_{(01)}^L&Q_{(12)}^L&Q_{(23)}^L&\ldots&Q_{(d-4\,d-3)}^L&Q_{(d-3\,d-2)}^L\\
\hline
X_1&-d&(d-2)&0&\ldots&\ldots&0\\
X_2&0&-(d-1)&(d-3)&\ldots&\ldots&0\\
X_3&0&0&-(d-2)&\ddots&\ldots&0\\
\vdots&\vdots&\vdots&\ddots&\ddots&\ddots&\vdots\\
X_{d-3}&0&\ldots&\ldots&0&-4&2\\
X_{d-2}&0&\ldots&\ldots&0&0&-3\\
\end{array}
\end{equation}
Namely, one obtains that ${\mathcal T}_{N_{(01)},\ldots,N_{(d-3\,d-2)}}$ is generated by generators 
whose charges satisfy 
\begin{equation}
\begin{array}{ccccc}
N_{(01)}-d&<&Q_{(01)}^L&\leq& N_{(01)}\\
N_{(12)}-(d-1)&<&Q_{(12)}^L&\leq& N_{(12)}\\
&&\vdots&&\\
N_{(d-3\,d-2)}-3&<&Q_{(d-3\,d-2)}^L&\leq& N_{(d-3\,d-2)}\\
\end{array}
\end{equation}
Indeed, these inequalities can be solved successively. The inequality for $Q_{(01)}^L$ can be rewritten as
\begin{equation}
-d<Q_{(01)}^L-N_{(01)}=Q_0^L-N_{(01)}\leq 0
\end{equation}
which has solutions
\begin{equation}
Q_0^L=N_{(01)}-a\,,\quad a\in\{0,\ldots,d-1\}.
\end{equation}
The inequality for $Q_{(12)}^L$ can be brought into the form
\begin{equation}
(d-1)(Q_1^L-1)<Q_0^L-N_{(12)}\leq(d-1)Q_1^L.
\end{equation}
Hence $Q_1^L$ is determined by $Q_0^L$:
\begin{equation}
Q_1^L=\left\lceil\frac{Q_0^L-N_{(12)}}{d-1}\right\rceil.
\end{equation}
In a similar way, the inequality for $Q_{(23)}^L$ determines $Q_2^L$ in terms of $Q_0^L$ and $Q_1^L$:
\begin{equation}
Q_2^L=\left\lceil\frac{Q_0^L-(d-1)Q_1^L-N_{(23)}}{d-2}\right\rceil.
\end{equation}
Going all the way we find $d$ solutions for the inequalities given by
\begin{equation}
\begin{array}{ll}
Q_0^L=N_{(01)}-a=:Q_0^L(a)\,,\quad a\in\{0,\ldots,d-1\}\\
Q_i=\left\lceil
\frac{Q_0^L-\sum_{j=1}^{i-1}(d-j)Q_j^L -N_{(i\,i+1)}}{d-i}
\right\rceil
=\left\lceil
\frac{N_{(01)}-N_{(i\,i+1)}-a-\sum_{j=1}^{i-1}(d-j)Q_j^L}{d-i}
\right\rceil=:Q_i^L(a)
\,,\;\text{for}\,i>0
\end{array}
\end{equation}
Thus, the truncated module has $d$ generators
\begin{equation}
e_{a}:=\alpha_0^{Q_0^L(a)}\cdot\ldots\cdot \alpha_{d-3}^{Q_{d-3}^L(a)},\;a\in\{0,\ldots,d-1\}.
\end{equation}
These satisfy relations
\begin{equation}
X_0^\prime e_a=X_0^{m_0^a}\cdot\ldots\cdot X_{d-2}^{m_{d-2}^a} e_{a-1},
\end{equation}
where the $m_i^a$ can take values $0$ or $1$ such that $\sum_a m_i^a=(d-i)$. In particular, $m_0^a=1$ for all $a$, $m_1^a=1$ for all but one $a$, $m_2^a=1$ for all but two $a$'s etc. Which of the $m_i^a$ are zero is determined by the choice of $N_{(i\,i+1)}$ in a rather complicated manner depending on divisibility properties. In fact, ${\mathcal T}_{N_{(01)},\ldots,N_{(d-3\,d-2)}}$ is a free module divided by these relations. 

For concreteness we will present the solution for the choice $N_{(i\,i+1)}=0$ for all $i$. In this case
\begin{equation}
Q_i^L(a)=-\delta_{a,d-i}\,,\;\text{for}\;i>0.
\end{equation}
In particular, the generators read
\begin{equation}
e_{a}=\left\{\begin{array}{ll}
\alpha_0^{-a},&0\leq a\leq 2\\
\alpha_0^{-a}\alpha_{d-a}^{-1},&3\leq a\leq d-1
\end{array}
\right.
\end{equation}
They satisfy the relations
\begin{equation}
\begin{array}{ll}
X_0^\prime e_0=X_0e_{d-1}&\\
X_0^\prime e_1=X_0\cdot\ldots\cdot X_{d-2} e_0&\\
X_0^\prime e_{a}=X_0\cdot\ldots\cdot X_{d-a} e_{a-1}&\text{for}\;2\leq a\leq d-1
\end{array}
\end{equation}
and ${\mathcal T}_{N_{(01)},\ldots,N_{(d-3\,d-2)}}$ can be presented as the cokernel of the matrix
\begin{equation}
p_1=\left(\small\begin{array}{ccccccccc}
X_0^\prime&&&&&&& -X_0\\
-X_0\cdot\ldots X_{d-2}&X_0^\prime &&&&&&\\
&-X_0\cdot\ldots\cdot X_{d-2}&X_0^\prime &&&&&\\
&&-X_0\cdot\ldots\cdot X_{d-3}&X_0^\prime&&&&\\
&&&-X_0\cdot\ldots\cdot X_{d-4}&&&&\\
&&&&\ddots&\ddots&&&\\
&&&&&\ddots&X_0^\prime&\\
&&&&&& -X_0X_1&X_0^\prime
\end{array}
\right)
\end{equation}
which is one of the matrices of the matrix factorization representing the lift defect.
Setting all but the $X_j=1$ for $j\neq i$ one obtains from this the defect describing the transition between phase $0$ and phase $j$ along the chosen path. Again we find agreement with the flow defects constructed in \cite{Brunner:2007ur}.

\section*{Acknowledgements}
We thank Johanna Knapp and Ed Segal for useful discussions.
IB is supported by the Deutsche Forschungsgemeinschaft (DFG, German Research Foundation) under Germany's Excellence Strategy -- EXC-2094 -- 390783311 and the DFG grant  ID 17448. DR thanks the Simons Center for Geometry and Physics for its hospitality.

\printbibliography
\end{document}